\documentclass[%
 reprint,
 superscriptaddress,
 amsmath,amssymb,
 aps,
]{revtex4-2}

\usepackage{graphicx}
\usepackage{dcolumn}
\usepackage{bm}
\usepackage[hidelinks]{hyperref}
\usepackage[all]{hypcap}    

\usepackage[nolist]{glossaries} 
\newacronym{QETU}{QETU}{quantum eigenvalue transformation of unitary matrices}
\newacronym{QSVT}{QSVT}{quantum singular value transformation}
\newacronym{FSWAP}{FSWAP}{fermionic swap}
\glsdisablehyper 

\usepackage[capitalise]{cleveref} 

\usepackage{orcidlink}

\usepackage{tikz}
\usepackage{pgfplots}
\pgfplotsset{compat=newest}
\usetikzlibrary{quantikz2}
\usetikzlibrary{matrix}
\usetikzlibrary{intersections}
\usetikzlibrary{calc}
\usepgfplotslibrary{fillbetween}
\usepgfplotslibrary{groupplots}
\usepgfplotslibrary{colorbrewer}

\usepackage{siunitx}
\usepackage{dsfont}
\usepackage[caption=false]{subfig}

\newcommand*\circled[1]{\tikz[baseline=0,yshift=1mm]{
            \node[shape=circle,draw,inner sep=2pt, scale=0.6] (char) {#1};}}

\begin{document}

\preprint{APS/123-QED}

\title{Ground-State Preparation of the Fermi-Hubbard Model on a Quantum Computer with 2D Topology via Quantum Eigenvalue Transformation of Unitary Matrices}

\author{Thilo R. Müller\,\orcidlink{0009-0007-8569-9138}}
\email{thilo.mueller@tum.de}
\affiliation{%
Technical University of Munich, School of CIT, Department of Computer Science, Boltzmannstraße 3, 85748 Garching, Germany
}%
\author{Manuel Geiger\,\orcidlink{0000-0003-3514-8657}}%
\email{manuel.geiger@tum.de}
\affiliation{%
Technical University of Munich, School of CIT, Department of Computer Science, Boltzmannstraße 3, 85748 Garching, Germany
}%
\author{Christian B. Mendl\,\orcidlink{0000-0002-6386-0230}}
\email{christian.mendl@tum.de}
\affiliation{%
Technical University of Munich, School of CIT, Department of Computer Science, Boltzmannstraße 3, 85748 Garching, Germany
}%
\affiliation{%
Technical University of Munich, Institute for Advanced Study, Lichtenbergstraße 2a, 85748 Garching, Germany
}%

\date{\today}

\begin{abstract}
Quantum computing holds immense promise for simulating quantum systems, a critical task for advancing our understanding of complex quantum phenomena. One of the primary goals in this domain is to accurately approximate the ground state of quantum systems. The Fermi-Hubbard model, particularly, is of profound interest due to its implications for high-temperature superconductivity and strongly correlated electron systems. The quantum eigenvalue transformation of unitary matrices (QETU) algorithm offers a novel approach for ground state estimation by utilizing a controlled Hamiltonian time evolution operator, circumventing the resource-intensive block-encoding required by previous methods. In this work, we apply the QETU algorithm to the $2 \times 2$ Fermi-Hubbard model, presenting circuit simplifications tailored to the model and introducing a mapping to a 9-qubit grid-like hardware architecture inspired by fermionic swap networks. We investigate how the selection of a favorable hardware architecture can benefit the circuit construction. Additionally, we explore the feasibility of this method under the influence of noise, focusing on its robustness and practical applicability.
\end{abstract}

\maketitle


\section{\label{sec:intro}Introduction}
Simulating a (strongly correlated) target quantum system is considered one of the most promising and foundational applications of quantum computing. Often, the primary interest is in calculating the ground state of a quantum system.

The recently developed \gls{QETU} \cite{Dong2022} aims for this goal. This algorithm is based on the \gls{QSVT} \cite{Gilyen2019, Low2019, Martyn2021} framework, which transforms the eigenvalues of a Hermitian matrix according to a tailored polynomial. While \gls{QSVT} in its original formulation requires a resource-intensive block-encoding of the input Hamiltonian, \gls{QETU} utilizes a controlled Hamiltonian time evolution operator instead. Moreover, the authors of \cite{Dong2022} introduce a method to approximate the controlled time evolution operator of certain quantum spin Hamiltonians through a Suzuki-Trotter decomposition and commuting Pauli strings to design an algorithm using only single- and two-qubit gates.

Despite its advantages over previous ground-state estimation approaches, the cost of implementing \gls{QETU} may still exceed the capabilities of early fault-tolerant quantum devices where limited coherence times are expected to remain an issue. Therefore, a key objective when implementing algorithms like \gls{QETU} on real hardware is to keep the circuit depths as small as possible. One way to achieve this objective is to select a hardware platform that offers a native gate set aligned with the specific algorithm’s requirements \cite{Schuch2003, Abrams2020}.

In this work, we apply and refine the \gls{QETU} algorithm to calculate the ground state of the $2 \times 2$ Fermi-Hubbard model \cite{Hubbard1963}. 
Specifically, we focus on investigating how the algorithm can be efficiently implemented on realistic quantum hardware and how the implementation can benefit from an optimal hardware architecture, aiming to understand the practical challenges and resource requirements of such implementations.
We present further simplifications of the \gls{QETU} circuit for the two-dimensional (2D) Fermi-Hubbard model and introduce a mapping to the specific grid-like qubit hardware architecture that is inspired by fermionic swap networks \cite{Kivlichan2018}.
The selected hardware platform offers an ideal foundation for the efficient implementation of the $2 \times 2$ toy model because both qubit topology and native gate set favor the implementation of a fermionic swap network. 
Furthermore, we investigate the applicability of our implementation under the influence of noise effects.

\section{\label{sec:background}Background}

\subsection{\label{sec:fermi_hubbard}Fermi-Hubbard model}
The Fermi-Hubbard model is a quantum many-body model frequently used in condensed matter physics to describe the behavior of interacting fermions on a lattice.
In this model, only so-called onsite Coulomb interactions are allowed. The repulsive Coulomb forces are restricted in their range so that they only act between electrons at the same site.
The kinetic energy is represented by electrons hopping between neighboring lattice sites.

On a 2D lattice, the Fermi-Hubbard model is defined as
\begin{align}
    \label{eq:fh}
    H_{\text{FH}} &= u \sum_{j=1}^n n_{i,\uparrow} n_{i,\downarrow} -t \sum_{\langle i,j \rangle} \sum_{\sigma \in \{ \uparrow,\downarrow \} } \left( a^{\dagger}_{i,\sigma} a_{j,\sigma} + a^{\dagger}_{j,\sigma} a_{i,\sigma} \right),
\end{align}
where $a^{\dagger}_{i,\sigma}$ and $a_{i,\sigma}$ are the creation and annihilation operators for spin $\sigma \in \{ \uparrow,\downarrow \}$ on lattice site $i$, and $n_{i,\sigma} = a^{\dagger}_{i,\sigma} a_{i,\sigma}$ is the number operator. The first term is the local interaction term that adds the on-site Coulomb repulsion energy $u$ if the corresponding lattice site is occupied by two electrons.
The second term is the kinetic interaction term, where $t$ is the kinetic hopping energy. It sums over all neighboring lattice sites $\langle i,j \rangle$ and describes the concurrent destruction of an electron at site $i$ and the creation of an electron at site $j$.
Using the Jordan-Wigner transformation \cite{JordanWigner1928, Ortiz2001}, the fermionic operators of the Fermi-Hubbard model can be represented as qubits suitable for a quantum computer (see \cref{sec:jw_transform} for a detailed derivation).
\begin{align}
    H &= \frac{1}{4} u \sum_{j=1}^n Z_{j,\uparrow} Z_{j,\downarrow} \nonumber \\
    & - \frac{1}{2} t \sum_{\langle i,j \rangle} \sum_{\sigma \in \{ \uparrow,\downarrow \} } \left( X_{i,\sigma} X_{j,\sigma} + Y_{i,\sigma} Y_{j,\sigma} \right) Z_{i+1,\sigma} \hdots Z_{j-1,\sigma}.
    \label{eq:fh_jw}
\end{align}

\subsection{\label{sec:qetu}Quantum Eigenvalue Transformation of Unitary Matrices}
\begin{figure*}
    \centering
    \begin{quantikz}
        \lstick{\ket{0}} & \gate{e^{i \varphi_0 X}} & \gate[2]{V} & \gate{e^{i \varphi_1 X}} & \gate[2]{V^{\dagger}} & \qw \ \ldots\ & \gate[2]{V} & \gate{e^{i \varphi_1 X}} & \gate[2]{V^{\dagger}} & \gate{e^{i \varphi_0 X}} & [0.2cm] & \meter{} \\
        \lstick{\ket{\psi}} & \qw & \qw & \qw & \qw & \qw \ \ldots\ & \qw & \qw & \qw & \qw & & \setwiretype{n} & \lstick{$\frac{f(H) \ket{\psi}}{\lVert f(H) \ket{\psi} \rVert}$}
    \end{quantikz}
    \caption{Circuit of the \gls{QETU} algorithm using a controlled forward and backward time evolution operator, adapted from \cite{Dong2022}.}
    \label{fig:qetu}
\end{figure*}
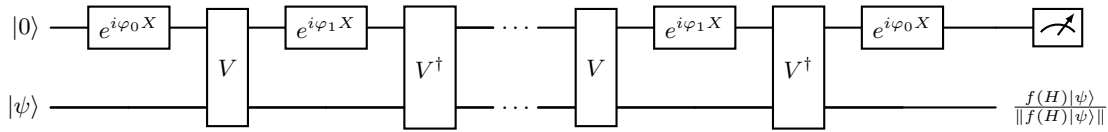

The \gls{QETU} algorithm enables the transformation of a Hermitian input matrix $H \in \mathbb{C}^{n \times n}$ according to a predefined target polynomial $F(x)$ of degree $d$ that must be real-valued, have even parity, and satisfy $\lvert F(x) \rvert \le 1$ for all $x \in [-1,1]$. This transformation can be seen as a matrix function, i.e., it transforms the eigenvalues of $H$.
The standard variation of \gls{QETU} requires a controlled version of the time evolution operator \begin{equation}
\label{eq:time_evolution_op}
    U(\Delta t) = e^{-i \Delta t H},
\end{equation}
where $H$ is the Hamiltonian of the system, $\Delta t$ is the evolution time, and we assume that $\hbar = 1$.
In this work, we use a special "control-free" variation of the \gls{QETU} algorithm defined in Corollary 17 of \cite{Dong2022} that assumes access to an oracle $V$ that implements a controlled forward and backward time evolution of the input matrix $H$ and is defined as
\begin{equation}
    V =
    \begin{pmatrix}
        e^{i \Delta t H} & 0 \\
        0 & e^{-i \Delta t H}
    \end{pmatrix}.
    \label{eq:forward_backward_time_evolution_op}
\end{equation}
The \gls{QETU} circuit, as visualized in \cref{fig:qetu}, is an alternating sequence of signal rotation gates that only act on the ancilla qubit and the oracle $V$. The polynomial $F(x)$ is encoded within the circuit through the parameters of the rotation gates that form a sequence of symmetric phase angles $\Phi_Z = (\phi_0, \phi_1, \dots, \phi_1, \phi_0) \in \mathbb{R}^{d+1}$. These phase angles can be determined using optimization-based methods to approximate the target polynomial \cite{Dong2021, Wang2022}. The QETU circuit then transforms the matrix $H$ as follows:
\begin{equation}
    \left( \bra{0} \otimes I_n \right) U_{\text{QETU}} \left( \ket{0} \otimes I_n \right) = F(\cos (H / 2))
\end{equation}

The operator $V$ is constructed by controlling the regular time evolution operator $U$ using Pauli strings as described in Section VI of \cite{Dong2022}.
Assume that the overall Hamiltonian can be grouped into $l$ terms of Pauli operators
\begin{equation}
    H = \sum_{j=1}^{l} H^{(j)}.
\end{equation}
Then, for each term $H^{(j)}$ in the Hamiltonian, we need to find a Pauli operator $K_j$ that anticommutes with the term to conjugate the time evolution operator, i.e., $K_j \cdot e^{-i H^{(j)} \Delta t} \cdot K_j = e^{i H^{(i)} \Delta t}$. Representing $V$ as a quantum circuit, we can see that only the Pauli operators are controlled by the ancilla qubit while the time evolution operator itself does not need to be controlled and can hence be efficiently approximated with the Trotter decomposition:
\begin{equation}
\begin{split}
    V &=
    \begin{pmatrix}
    K e^{-i \Delta t H} K & 0 \\
    0 & e^{-i \Delta t H}
    \end{pmatrix} \\
    &=
    \begin{quantikz}
        \qw & \octrl{1} & \qw & \octrl{1} & \qw \\
        \qw & \gate{K} & \gate{e^{-i \Delta t H}} & \gate{K} & \qw
    \end{quantikz}
\end{split}
\end{equation}

\subsection{Eigenspace filtering}
As mentioned in the beginning, the ground state plays a crucial role in understanding the fundamental properties of the system under investigation. Since the ground state corresponds to the eigenvector associated with the smallest eigenvalue of a Hamiltonian, the \gls{QETU} algorithm estimates the ground state by filtering out all other eigenvalues. Thus, under the assumption that we have a non-zero overlap $\gamma = \lvert \braket{\psi_0}{\psi_{\text{init}}} \rvert$ between the initial state and the ground state of the system, we need to find a function that amplifies only the ground state vector and diminishes all the other components.
Additionally, the Hamiltonian needs to be scaled such that its spectrum is in the range $\left[ \eta, \pi-\eta \right]$ for some constant $\eta > 0$. This is because the cosine function is periodic and thus, the filter function would repeat itself if the range were not limited.
Assume that $\Delta = \lambda_1 - \lambda_0$ denotes the spectral gap, i.e., the distance between the ground-state energy and the first exited-state energy. Then, as stated in Theorem 6 of \cite{Dong2022}, a suitable function to isolate the ground state is the shifted sign function
\begin{equation}
    g(x) = \frac{1}{2} \left( 1 - \text{sgn} (x-\mu) \right) = \begin{cases}
        1 & x \le \mu \\
        0 & x > \mu \\
    \end{cases},
\end{equation}
where
\begin{equation}
    \mu = \frac{1}{2} \left( \lambda_0 + \lambda_1 \right),
\end{equation}
such that
\begin{equation}
    \lambda_0 \le \mu - \Delta/2 \le \mu + \Delta/2 \le \lambda_1
\end{equation}
is satisfied. The shifted sign function can be approximated by the real polynomial $f(x)$ satisfying
\begin{subequations}
\begin{align}
    \lvert f(x)-c \rvert \le \epsilon \quad &\text{for} \, x \in \left[ \eta, \mu - \Delta/2 \right], \\
    \lvert f(x) \rvert \le \epsilon \quad &\text{for} \, x \in \left[ \mu + \Delta/2, \pi - \eta \right].
\end{align}
\end{subequations}
Here, the parameter $c$ is chosen to be slightly smaller than 1 to avoid numerical overshooting when finding the polynomial approximation to the constraints above. The maximally allowed error is denoted by $\epsilon$.
Taking into account the cosine transformation that occurs during the application of the \gls{QETU} algorithm, we need to find a polynomial $F(x)$ that satisfies
\begin{subequations}
\begin{align}
    \lvert F(x)-c \rvert \le \epsilon \quad &\text{for} \, x \in \left[ \sigma_{+}, \sigma_{\text{max}} \right], \\
    \lvert F(x) \rvert \le \epsilon \quad &\text{for} \, x \in \left[ \sigma_{\text{min}}, \sigma_{-} \right], \\
    \lvert F(x) \rvert \le 1 \quad &\text{for} \, x \in \left[ -1, 1 \right],
\end{align}
\end{subequations}
where
\begin{subequations}
\begin{align}
    \sigma_{\pm} &= \cos \left( \mu \mp \Delta/2 \right), \\
    \sigma_{\text{min}} &= \cos \left( \pi - \eta \right), \\
    \sigma_{\text{max}} &= \cos \eta.
\end{align}
\end{subequations}

\begin{figure}
    \centering
    \resizebox{\columnwidth}{!}{\input{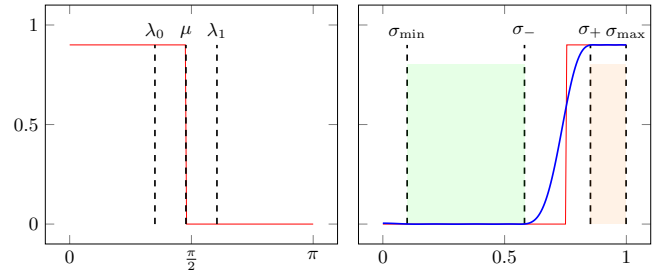}}
    \caption{Left: Plot of the step function $g(x)$ that can be used to filter out all eigenvalues except the ground-state energy. The value $\mu$ is chosen to be halfway between the ground-state energy $\lambda_0$ and the first exited-state energy $\lambda_1$. Right: The red curve $g(2 \arccos x)$ shows the step function from the left graph with a transformed domain to account for the cosine transformation of the \gls{QETU} algorithm. The blue curve shows $F(x)$ which is the polynomial approximation of degree $d=30$ of the step function. The dashed black lines indicate the values $\sigma_{\text{min}}$, $\sigma_{\pm}$, and $\sigma_{\text{max}}$. The orange colored area indicates the interval $\left[ \sigma_{+}, \sigma_{\text{max}} \right]$, while the green shaded area indicates the interval $\left[ \sigma_{\text{min}}, \sigma_{-} \right]$.}
    \label{fig:step_function}
\end{figure}

The left plot of \cref{fig:step_function} shows an example of the step function $g(x)$ for parameters $\mu = 1.5$, $\lambda_0 = 1.1$, and $\lambda_1 = 1.9$. The right plot shows the composition of the step function and an argument transformation $2 \arccos x$ that is necessary to achieve the effect of the regular step function when using the \gls{QETU} circuit. Moreover, the plot shows the polynomial approximation $F(x)$ to that transformed step function.

\subsection{Fermionic swap networks}
The Fermi-Hubbard model describes interacting fermions on a lattice, which involves fermionic creation and annihilation operators that obey certain anti-commutation relations. Quantum computers, however, are typically based on qubits and use operators that satisfy the standard Pauli algebra. Since fermionic operators and Pauli operators represent fundamentally different types of quantum systems and have distinct algebraic properties, the challenge is to map fermionic operators to qubit operators while preserving the essential properties of the original system.

Various techniques exist to achieve this mapping, most notably the Jordan-Wigner transformation.
While the Jordan-Wigner transformation is a valuable tool for simulating fermionic systems on quantum computers, it does have certain limitations and drawbacks, especially when applied to complex systems like the 2D Fermi-Hubbard model.
Most importantly, the Jordan-Wigner transformation can lead to the creation of $Z$-strings in the resulting qubit Hamiltonian. These $Z$-strings correspond to non-local terms that connect qubits across different lattice sites in the original fermionic system. The presence of $Z$-strings creates entanglement between non-neighboring qubits, necessitating long-range interactions in quantum hardware. These long-range interactions can be challenging to implement and result in additional resource overhead, as it is necessary to decompose them into hardware-native operations and insert additional $\operatorname{SWAP}$-gates so that the respective gates can be applied on physically adjacent qubits \cite{Wille2014}. This leads to an increased gate count and circuit depth, particularly on devices with limited qubit connectivity, and consequently, exacerbates the effects of noise and decoherence.

\begin{figure}
\subfloat[\label{subfig:a}]{%
  \resizebox{0.45\columnwidth}{!}{\begin{tikzpicture}[
    scale=0.65,
    transform shape,
    darkstyle/.style={circle,draw,fill=gray!40,minimum size=25}, bluestyle/.style={circle,draw,fill=blue!40,minimum size=25},
    redstyle/.style={circle,draw,fill=red!40,minimum size=25},
    greenstyle/.style={circle,draw,fill=gray!20,minimum size=25}]
        \node [redstyle]  (02) at (0*1.5,2*1.5) {$1 \uparrow$};
        \node [bluestyle]  (12) at (1*1.5,2*1.5) {$1 \downarrow$};
        \node [bluestyle]  (22) at (2*1.5,2*1.5) {$2 \downarrow$};
        \node [redstyle]  (32) at (3*1.5,2*1.5) {$2 \uparrow$};
        \node [redstyle]  (01) at (0*1.5,1*1.5) {$3 \uparrow$};
        \node [bluestyle]  (11) at (1*1.5,1*1.5) {$3 \downarrow$};
        \node [bluestyle]  (21) at (2*1.5,1*1.5) {$4 \downarrow$};
        \node [redstyle]  (31) at (3*1.5,1*1.5) {$4 \uparrow$};

        \draw (02) edge[->] node[font=\small\ttfamily,above,xshift=-2pt] {1} (12);
        \draw (12) edge[->] node[font=\small\ttfamily,above,xshift=-2pt] {2} (22);
        \draw (22) edge[->] node[font=\small\ttfamily,above,xshift=-2pt] {3} (32);
        \draw (32) edge[->] node[font=\small\ttfamily,right,yshift=2pt] {4} (31);
        \draw (31) edge[->] node[font=\small\ttfamily,above,xshift=2pt] {5} (21);
        \draw (21) edge[->] node[font=\small\ttfamily,above,xshift=2pt] {6} (11);
        \draw (11) edge[->] node[font=\small\ttfamily,above,xshift=2pt] {7} (01);
        \draw (01) edge[->] node[font=\small\ttfamily,left,yshift=-2pt] {8} (02);

\end{tikzpicture}}
}\hfill
\subfloat[\label{subfig:b}]{%
  \resizebox{0.45\columnwidth}{!}{\begin{tikzpicture}[
    scale=0.65,
    transform shape,
    darkstyle/.style={circle,draw,fill=gray!40,minimum size=25}, bluestyle/.style={circle,draw,fill=blue!40,minimum size=25},
    redstyle/.style={circle,draw,fill=red!40,minimum size=25},
    greenstyle/.style={circle,draw,fill=gray!20,minimum size=25}]
        \node [redstyle]  (02) at (0*1.5,2*1.5) {$1 \uparrow$};
        \node [bluestyle]  (12) at (1*1.5,2*1.5) {$1 \downarrow$};
        \node [bluestyle]  (22) at (2*1.5,2*1.5) {$2 \downarrow$};
        \node [redstyle]  (32) at (3*1.5,2*1.5) {$2 \uparrow$};
        \node [redstyle]  (01) at (0*1.5,1*1.5) {$3 \uparrow$};
        \node [bluestyle]  (11) at (1*1.5,1*1.5) {$3 \downarrow$};
        \node [bluestyle]  (21) at (2*1.5,1*1.5) {$4 \downarrow$};
        \node [redstyle]  (31) at (3*1.5,1*1.5) {$4 \uparrow$};

        \draw (02) edge[latex reversed-latex reversed] node[font=\tiny\ttfamily,above] {} (12);
        \draw (22) edge[latex reversed-latex reversed] node[font=\tiny\ttfamily,above] {} (32);
        \draw (31) edge[latex reversed-latex reversed] node[font=\tiny\ttfamily,above] {} (21);
        \draw (11) edge[latex reversed-latex reversed] node[font=\tiny\ttfamily,above] {} (01);

\end{tikzpicture}}
}\\
\subfloat[\label{subfig:c}]{%
  \resizebox{0.45\columnwidth}{!}{\begin{tikzpicture}[
    scale=0.65,
    transform shape,
    darkstyle/.style={circle,draw,fill=gray!40,minimum size=25}, bluestyle/.style={circle,draw,fill=blue!40,minimum size=25},
    redstyle/.style={circle,draw,fill=red!40,minimum size=25},
    greenstyle/.style={circle,draw,fill=gray!20,minimum size=25}]
        \node [redstyle]  (02) at (0*1.5,2*1.5) {$1 \uparrow$};
        \node [bluestyle]  (12) at (1*1.5,2*1.5) {$1 \downarrow$};
        \node [bluestyle]  (22) at (2*1.5,2*1.5) {$2 \downarrow$};
        \node [redstyle]  (32) at (3*1.5,2*1.5) {$2 \uparrow$};
        \node [redstyle]  (01) at (0*1.5,1*1.5) {$3 \uparrow$};
        \node [bluestyle]  (11) at (1*1.5,1*1.5) {$3 \downarrow$};
        \node [bluestyle]  (21) at (2*1.5,1*1.5) {$4 \downarrow$};
        \node [redstyle]  (31) at (3*1.5,1*1.5) {$4 \uparrow$};

        \draw (12) edge[latex reversed-latex reversed] node[font=\tiny\ttfamily,above] {} (22);
        \draw (32) edge[latex reversed-latex reversed] node[font=\tiny\ttfamily,above] {} (31);
        \draw (21) edge[latex reversed-latex reversed] node[font=\tiny\ttfamily,above] {} (11);
        \draw (01) edge[latex reversed-latex reversed] node[font=\tiny\ttfamily,above] {} (02);

\end{tikzpicture}}
}
\caption{(a) Depiction of the mapping of the $2 \times 2$ Fermi-Hubbard model to a one-dimensional loop of qubits. Each circle represents a spin-orbital while the arrows between the circles indicate the canonical ordering of the Jordan-Wigner transformation. The path is then decomposed into a left stagger (b), containing all the odd edges, and a right stagger (c), which contains all the even edges.}\label{fig:qubit_chain}
\label{fig:linear_qubit_chain}
\end{figure}
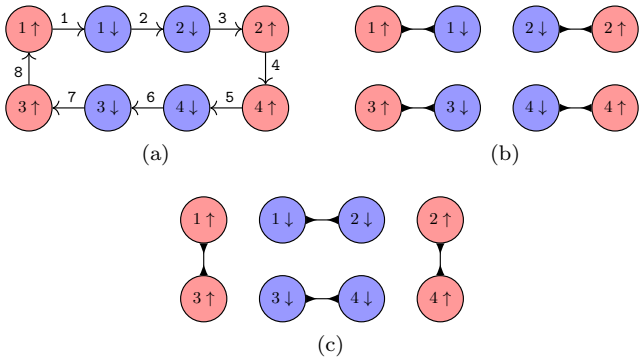

Fermionic swap networks provide an elegant solution to address the overhead of the non-local interactions introduced by the Jordan-Wigner transformation. In fermionic swap networks, the mapping between physical qubits and fermionic modes is dynamically changed to bring interacting fermionic modes into adjacency and implement the interaction locally using fermionic swap gates. Each swap gate simultaneously performs a swap operation while preserving the anti-symmetry and evolving the respective qubit pair under the kinetic hopping or onsite-interaction operator.
The optimal sequence of fermionic swaps is determined by defining a linear path through the qubits and decomposing it into a "left stagger" ($U_L$) and a "right stagger" ($U_R$). Following that qubit path, the connections between the qubits are alternately assigned to $U_L$ or $U_R$. Each connection is then considered a fermionic swap operation that can be applied in parallel alongside the other operations in the same stagger. Then, by alternately applying $U_L$ and $U_R$, each spin-orbital will eventually become adjacent to all others at one point in the circuit. A visualization of this construction for the $2 \times 2$ Fermi-Hubbard model is given in \cref{fig:qubit_chain}.
In the case of the 2D Fermi-Hubbard model, the ordering of the spin-orbitals follows the canonical ordering of fermions \cite{Somma2002} where the 2D lattice is transversed in a zigzag pattern.
Moreover, the implementation of the fermionic swap network depends crucially on the underlying physical qubit layout. In a grid-like qubit layout, the physical arrangement is more closely aligned with the structure of the 2D Fermi-Hubbard model, reducing the number of fermionic swaps required to implement fermionic interactions.

In this work, we mapped the spin-orbitals onto the grid-like hardware according to the canonical ordering but included the auxiliary qubit in the center of the grid. With that arrangement, and due to the small size of the system, only two distinct mapping configurations are required to bring all relevant orbitals into an adjacent position at least once in the circuit. Moreover, the auxiliary qubit becomes adjacent to every other spin-orbital at least once, enabling the implementation of the controlled Pauli operators based on two-qubit gates. The fermionic simulation gate used in the original paper is not natively available on the quantum hardware used in this work. Instead, onsite interaction, kinetic hopping, and fermionic swap could each be implemented with a single two-qubit gate. Consequently, the onsite interaction and parts of the kinetic hopping were carried out on the same qubit arrangement but between complementary qubit pairs. This approach is more efficient than performing all cycles necessary for a complete rotation of qubits, as outlined in the original fermionic swap network.

\section{\label{sec:hardware}Hardware platform}
The target device used for the experiments is a superconducting quantum computer with 9 qubits arranged on a $3 \times 3$ grid topology. \cref{table:gate_set} lists the gates that are natively available on the device. Arbitrary $R_z$ rotations can be implemented virtually with zero cost by adding a phase offset to subsequent gates (see \cite{McKay2017} for details).

\begin{table}[h]
\footnotesize
    \caption{Native gate set of the quantum computer used for the experiments.}
    \label{table:gate_set}
    \begin{ruledtabular}
        \begin{tabular}{lcc}
            Name & Circuit diagram & Matrix \\
            \colrule
            Pauli X & \begin{quantikz}
                \qw & \gate{X} & \qw
            \end{quantikz} & $\begin{pmatrix}
                0 & 1 \\
                1 & 0
            \end{pmatrix}$ \\[10pt]
            Pauli Y & \begin{quantikz}
                \qw & \gate{Y} & \qw
            \end{quantikz} & $\begin{pmatrix}
                0 & -i \\
                i & 0
            \end{pmatrix}$ \\[10pt]
            $R_z(\lambda)$ & \begin{quantikz}
                \qw & \gate{R_z(\lambda)} & \qw
            \end{quantikz} & $\begin{pmatrix}
                e^{-i\frac{\lambda}{2}} & 0 \\
                0 & e^{i\frac{\lambda}{2}}
            \end{pmatrix}$ \\[10pt]
            $\sqrt{X}$ & \begin{quantikz}
                \qw & \gate{\sqrt{X}} & \qw
            \end{quantikz} & $\frac{1}{2} \begin{pmatrix}
                1+i & 1-i \\
                1-i & 1+i
            \end{pmatrix}$ \\[10pt]
            $\sqrt{Y}$ & \begin{quantikz}
                \qw & \gate{\sqrt{Y}} & \qw
            \end{quantikz} & $\frac{1}{2} \begin{pmatrix}
                1+i & -1-i \\
                1+i & 1+i
            \end{pmatrix}$ \\[10pt]
            CPhase($\lambda$) & \begin{quantikz}
                \qw & \ctrl{1} & \qw \\
                \qw & \gate{P(\lambda)} & \qw
            \end{quantikz} & $\begin{pmatrix}
                1 & 0 & 0 & 0 \\
                0 & 1 & 0 & 0 \\
                0 & 0 & 1 & 0 \\
                0 & 0 & 0 & e^{i\lambda}
            \end{pmatrix}$ \\[10pt]
            iSWAP($\theta$, $\eta$) & \begin{quantikz}
                \qw & \push{\bigotimes} \wire[d][1]["(\theta ; \, \eta)"]{q} & \qw \\
	           \qw & \push{\bigotimes} & \qw
            \end{quantikz} & $\begin{pmatrix}
                1 & 0 & 0 & 0 \\
                0 & \cos(\frac{\theta}{2}) & i e^{i\eta} \sin(\frac{\theta}{2}) & 0 \\
                0 & i e^{-i\eta} \sin(\frac{\theta}{2}) & \cos(\frac{\theta}{2}) & 0 \\
                0 & 0 & 0 & 1
            \end{pmatrix}$
        \end{tabular}
    \end{ruledtabular}
\end{table}

\section{\label{sec:time_evolution}Implementation of the Time Evolution Operator on a Quantum Device}
In this section, we describe the concrete implementation of the controlled forward and backward time evolution operator $V$ on the quantum computing architecture described in \cref{sec:hardware} for the $2 \times 2$ Fermi-Hubbard model.
First, we explain the mapping of the spin-states onto the physical qubits in \cref{sec:mapping}. The spin-states were placed in such a way as to facilitate the implementation of a fermionic swap network and thereby make optimal use of the qubit topology and connectivity on the available quantum device. This process is inherently connected with encoding the fermionic modes as qubits using the Jordan-Wigner transformation because vertically adjacent sites in the physical Fermi-Hubbard model translate to additional $Z$-strings between the respective qubits in the Jordan-Wigner representation. Hence, positioning the spin-states onto the qubit layout influences the Jordan-Wigner representation and, therefore, also the Trotter decomposition.
Second, we address the implementation of the time-evolution operator $U$ and how it is approximated using the Trotter formula in \cref{sec:trotterization}.
In \cref{sec:physical_realization}, we then present how we implemented the Jordan-Wigner transformed Hamiltonian on the target device, which is followed by \cref{sec:pauli_control}, where we describe how we control the sign of the resulting time-evolution operator $U$ to end up with the controlled forward and backward time-evolution operator $V$.
Lastly, we describe the details of rescaling the spectrum of the Hamiltonian in \cref{sec:hamiltonian_shift}.

\begin{figure*}
\includegraphics[width=2\columnwidth]{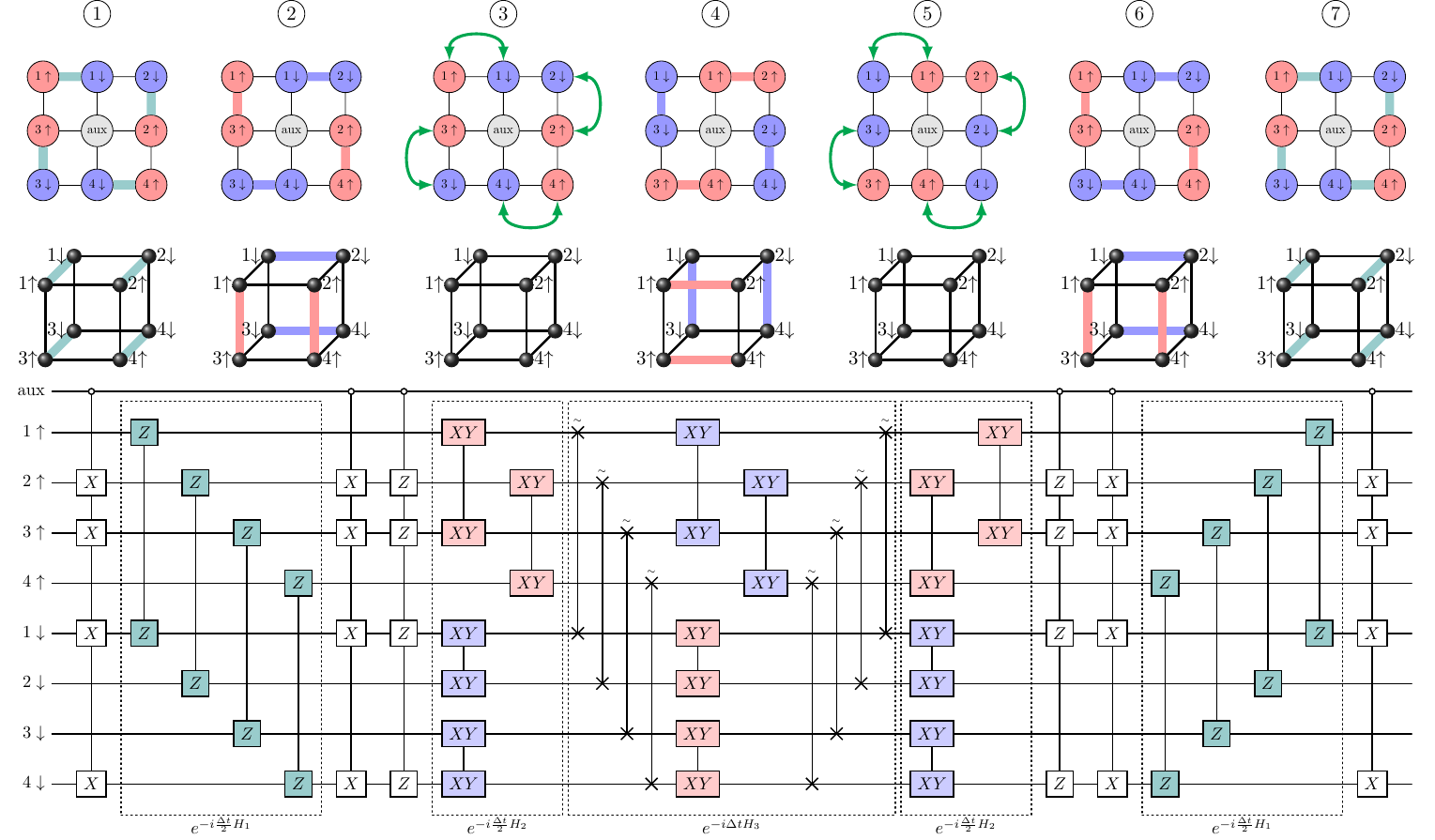}
\caption{\label{fig:time_evolution_op}A visualization of the quantum circuit that realizes one step of the second-order Trotter decomposition of the controlled forward and backward time evolution operator $V$ and the respective qubit arrangements for each part. The top part of the figure shows the mappings of the spin orbitals onto the physical hardware layout. Here, spin-up states are denoted with red, and spin-down states with blue. The colored edges symbolize that a gate is applied to the connected qubits. As can be seen, gates are only applied between neighboring qubits thanks to the fermionic swap operators which are depicted by the green arrows. The interaction term is realized in the start configuration \protect\circled{1}; the first part of the hopping term corresponds to configuration \protect\circled{2}. The second part of the hopping term is realized by first swapping the spin orbitals into neighboring positions in configuration \protect\circled{3}, then applying the hopping operators in \protect\circled{4}, and finally swapping them back into their original position in \protect\circled{5}. Due to the symmetry of the Trotter step, the second part of the hopping term and the interaction term are again applied in configurations \protect\circled{6} and \protect\circled{7}, respectively.}
\end{figure*}

\subsection{Mapping the physical model to qubits}
\label{sec:mapping}
As previously mentioned, the objective of the mapping is to place the spin states onto the physical qubits in a way that minimizes the number of additional operations needed to capture the canonical anti-commutation relations of the fermions in the original model. A variation of the fermionic swap network scheme can be used to find an optimal solution to this objective. For that, we split the original Hamiltonian into three smaller terms that are easier to deal with by itself.
Explicitly, in fermionic operator form, these are given by:
\begin{equation}
    H_1 = u \left( n_{1,\uparrow} n_{1,\downarrow} +  n_{2,\uparrow} n_{2,\downarrow} + n_{3,\uparrow} n_{3,\downarrow} + n_{4,\uparrow} n_{4,\downarrow} \right),
\end{equation}
which is exactly the on-site interaction term. The kinetic hopping term is split into two parts.
\begin{align}
    H_2 = -t \Bigl(
    \left( a^{\dagger}_{1,\uparrow} a_{3,\uparrow} + a^{\dagger}_{3,\uparrow} a_{1,\uparrow} \right)
    + \left( a^{\dagger}_{2,\uparrow} a_{4,\uparrow} + a^{\dagger}_{4,\uparrow} a_{2,\uparrow} \right)
    \nonumber \\
    + \left( a^{\dagger}_{1,\downarrow} a_{2,\downarrow} + a^{\dagger}_{2,\downarrow} a_{1,\downarrow} \right)
    + \left( a^{\dagger}_{3,\downarrow} a_{4,\downarrow} + a^{\dagger}_{3,\downarrow} a_{4,\downarrow} \right)
    \Bigr),
\end{align}
accounts for hopping in the vertical direction for spin-up fermions and the horizontal direction for spin-down fermions.
\begin{align}
    H_3 = -t \Bigl(
    \left( a^{\dagger}_{1,\downarrow} a_{3,\downarrow} + a^{\dagger}_{3,\downarrow} a_{1,\downarrow} \right)
    + \left( a^{\dagger}_{2,\downarrow} a_{4,\downarrow} + a^{\dagger}_{4,\downarrow} a_{2,\downarrow} \right)
    \nonumber \\
    + \left( a^{\dagger}_{1,\uparrow} a_{2,\uparrow} + a^{\dagger}_{2,\uparrow} a_{1,\uparrow} \right)
    + \left( a^{\dagger}_{3,\uparrow} a_{4,\uparrow} + a^{\dagger}_{3,\uparrow} a_{4,\uparrow} \right)
    \Bigr),
\end{align}
describes the hopping of fermions between adjacent sites but in complementary directions.

By choosing the initial fermion-to-qubit mapping as depicted in \cref{fig:time_evolution_op}, we can immediately simulate the interaction term $H_1$ using only local two-qubit gates. The general pattern is to place spin states corresponding to the same local site adjacent to each other along an S-pattern and alternate the order of spin-up and spin-down states. Note that in the $2 \times 2$ case, placing the auxiliary qubit in the center of the grid ensures that it is adjacent to one of the two possible spin states associated with each fermion.
The resulting Jordan-Wigner transformed interaction term is
\begin{align}
    H_1 = \frac{1}{4} u \bigl( &Z_{1,\uparrow} Z_{1,\downarrow} + Z_{2,\uparrow} Z_{2,\downarrow} \nonumber \\
    &+ Z_{3,\uparrow} Z_{3,\downarrow} + Z_{4,\uparrow} Z_{4,\downarrow} \bigr).
\end{align}

We can also directly implement the first half of the hopping term on the same qubit arrangement, namely, all the vertical interactions between spin-up states and all horizontal interactions between spin-down states. Under the Jordan-Wigner transformation, this part of the hopping term becomes
\begin{align}
    H_2 = -\frac{1}{2} t \bigl( &X_{1,\uparrow} X_{3,\uparrow} + Y_{1,\uparrow} Y_{3,\uparrow} \nonumber \\
    &+ X_{2,\uparrow} X_{4,\uparrow} + Y_{2,\uparrow} Y_{4,\uparrow} \nonumber \\
    &+ X_{1,\downarrow} X_{2,\downarrow} + Y_{1,\downarrow} Y_{2,\downarrow} \nonumber \\
    &+ X_{3,\downarrow} X_{4,\downarrow} + Y_{3,\downarrow} Y_{4,\downarrow} \bigr)].
\end{align}

To simulate the other half of the hopping term, we rearrange the fermionic modes such that all the horizontal spin-up states and vertical spin-down states are neighboring sites in the qubit layout.
The resulting term, however, contains additional Z-strings between the spin-up and spin-down states as they are not adjacent on the initial qubit layout. 
\begin{align}
    H_3 = -\frac{1}{2} t \bigl( &X_{1,\uparrow} X_{2,\uparrow} Z_{1,\downarrow} Z_{2,\downarrow} + Y_{1,\uparrow} Y_{2,\downarrow} Z_{1,\downarrow} Z_{2,\downarrow} \nonumber \\
    &+ X_{3,\uparrow} X_{4,\uparrow} Z_{3,\downarrow} Z_{4,\downarrow} + Y_{3,\uparrow} Y_{4,\uparrow} Z_{3,\downarrow} Z_{4,\downarrow} \nonumber \\
    &+ Z_{1,\uparrow} Z_{3,\uparrow} X_{1,\downarrow} X_{3,\downarrow} + Z_{1,\uparrow} Z_{3,\uparrow} Y_{1,\downarrow} Y_{3,\downarrow} \nonumber \\
    & + Z_{2,\uparrow} Z_{4,\uparrow} X_{2,\downarrow} X_{4,\downarrow} + Z_{2,\uparrow} Z_{4,\uparrow} Y_{2,\downarrow} Y_{4,\downarrow} \bigr).
\end{align}

\subsection{Trotter Splitting}
\label{sec:trotterization}
Realizing the entire time evolution operator becomes infeasible with increasing system sizes. The Trotter decomposition allows us to approximate the time evolution of a complex quantum system as a product of simpler one- and two-qubit gates. 
As mentioned in the previous section, the overall Hamiltonian can be grouped into three non-commuting terms $H = H_{1} + H_{2} + H_{3}$, which are simple enough to be implemented on the hardware as will be shown in the next section.

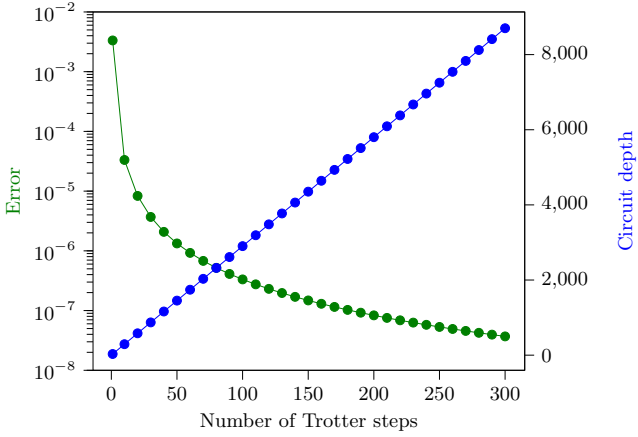
\begin{figure}
    \centering
    \resizebox{\columnwidth}{!}{
\begin{tikzpicture}

\definecolor{darkgray176}{RGB}{176,176,176}
\definecolor{green01270}{RGB}{0,127,0}

\begin{semilogyaxis}[
tick align=outside,
tick pos=left,
x grid style={darkgray176},
xlabel={Number of Trotter steps},
xmin=-13.95, xmax=314.95,
ylabel=\textcolor{green01270}{Error},
x label style={at={(0.5,-0.1)}},
ymin=1e-08, ymax=1e-02,
ymode=log,
ytick style={color=black},
ytick={1e-08, 1e-07, 1e-06, 1e-05,0.0001,0.001,0.01,0.1,1,10,100},
yticklabels={
  \(\displaystyle {10^{-8}}\),
  \(\displaystyle {10^{-7}}\),
  \(\displaystyle {10^{-6}}\),
  \(\displaystyle {10^{-5}}\),
  \(\displaystyle {10^{-4}}\),
  \(\displaystyle {10^{-3}}\),
  \(\displaystyle {10^{-2}}\),
  \(\displaystyle {10^{-1}}\),
  \(\displaystyle {10^{0}}\),
  \(\displaystyle {10^{1}}\),
  \(\displaystyle {10^{2}}\)
},
xtick style={draw=none},
xmajorticks=false,
]
\addplot [green01270, mark=*]
table {%
1 0.003334568714409329
10 3.326364504132173e-05
20 8.315755936097726e-06
30 3.6958787435111252e-06
40 2.0789292783088738e-06
50 1.330513992990914e-06
60 9.239677695525321e-07
70 6.788333390355751e-07
80 5.197317142467199e-07
90 4.106521846736645e-07
100 3.326282494410566e-07
110 2.748993704250505e-07
120 2.309918235455693e-07
130 1.96821430152825e-07
140 1.69708271312059e-07
150 1.4783476019718281e-07
160 1.2993289241117333e-07
170 1.1509626157373248e-07
180 1.0266302379160448e-07
190 9.214077672403066e-08
200 8.315704809179369e-08
210 7.542589602967693e-08
220 6.872483065728671e-08
230 6.287867381582175e-08
240 5.7747947431730385e-08
250 5.322050951803624e-08
260 4.920535503935993e-08
270 4.562800884876661e-08
280 4.2427065736469794e-08
290 3.95515096149665e-08
300 3.695868517164311e-08
};
\end{semilogyaxis}

\begin{axis}[
axis y line=right,
tick align=outside,
x grid style={darkgray176},
xmin=-13.95, xmax=314.95,
xtick pos=left,
xtick style={color=black},
y grid style={darkgray176},
ylabel=\textcolor{blue}{Circuit depth},
ymin=-402.55, ymax=9135.55,
ytick pos=right,
ytick style={color=black},
y label style={at={(1.2,0.5)}},
axis line style={-}
]
\addplot [blue, mark=*]
table {%
1 31
10 292
20 582
30 872
40 1162
50 1452
60 1742
70 2032
80 2322
90 2612
100 2902
110 3192
120 3482
130 3772
140 4062
150 4352
160 4642
170 4932
180 5222
190 5512
200 5802
210 6092
220 6382
230 6672
240 6962
250 7252
260 7542
270 7832
280 8122
290 8412
300 8702
};
\end{axis}

\end{tikzpicture}}
    \caption{The convergence of the Trotterization and the depth of the respective circuit on the target device using only native gates. The green line shows the $L_2$-norm error between the controlled forward and backward time evolution operator $V$ and its Trotter approximation on a logarithmic scale. The blue line shows the depth of the respective circuit without counting $R_z$-rotations.}
    \label{fig:trotter_error}
\end{figure}

Using the symmetric second-order Trotter-Suzuki decomposition of three parts (Eq. (62) of \cite{Hatano2005}), we can break down the time evolution operator into smaller steps that are feasible to implement on a quantum computer like
\begin{align}
    &e^{-i \Delta t H} \nonumber \\
    &= e^{-i \Delta t (H_{1} + H_{2} + H_{3})} \nonumber \\
    &= \left[ S_2(\Delta t) \right]^n + \mathcal{O}(\Delta t^3),
\end{align}
where
\begin{equation}
    S_2(\Delta t) = e^{-i \frac{\Delta t}{2n} H_{1}} e^{-i \frac{\Delta t}{2n} H_{2}} e^{-i \frac{\Delta t}{n} H_{3}} e^{-i \frac{\Delta t}{2n} t H_{2}} e^{-i \frac{\Delta t}{2n} H_{1}}
\end{equation}
is one Trotter step and $n$ is the total number of Trotter steps. Increasing the number of Trotter steps leads to a more accurate approximation, however, it also increases the number of quantum gates. This trade-off between precision and required hardware resources is depicted in \cref{fig:trotter_error}.  
The quantum circuit for one Trotter step is depicted in \cref{fig:time_evolution_op}. Since the first and the last part of the Trotter step (i.e., $\exp \left[ -i \frac{\Delta t}{2n} H_1 \right]$ and the Pauli string to control the sign of the time evolution) are identical, they can be merged together when repeating the Trotter steps to obtain the overall Trotter approximation.

\subsection{Physical realization}
\label{sec:physical_realization}
We can implement the whole interaction term $H_1$ by decomposing it into a product of $R_{zz}$ gates
\begin{align}
    e^{-i \frac{\Delta t}{2} H_1 / n}
    &= \exp \left[ -i \frac{\Delta t}{2n} \cdot \frac{1}{4} u \sum_{j=1}^n Z_{j,\uparrow} Z_{j,\downarrow} \right] \nonumber \\
    &= \prod_{j=1}^n \exp \left[ -i \frac{\Delta t}{2n} \cdot \frac{1}{4} u \, Z_{j,\uparrow} Z_{j,\downarrow} \right] \nonumber \\
    &= \prod_{j=1}^n R_{zz} \left( \frac{u \Delta t}{4n}\right),
\end{align}
which in turn can be implemented using $R_z$ and CPhase gates (see \cref{sec:gate_decomposition}).

The fist part of the kinetic hopping term only ever acts on neighboring qubits, and we can hence decompose it into
\begin{align}
    &e^{-i \frac{\Delta t}{2} H_2 / n} \nonumber \\
    &= \exp \left[ -i \frac{\Delta t}{2n} \cdot \left(-\frac{1}{2}t\right) \sum_{\langle i,j \rangle} \sum_{\sigma \in \{\uparrow,\downarrow\}} \left( X_{i,\sigma} X_{j,\sigma} + Y_{j,\sigma} Y_{i,\sigma} \right) \right] \nonumber \\
    &= \prod_{\langle i,j \rangle,\sigma} \exp \left[ -i \frac{(-t) \Delta t}{4n} \left( X_{i,\sigma} X_{j,\sigma} + Y_{i,\sigma} Y_{j,\sigma} \right) \right] \nonumber \\
     &= \prod_{\langle i,j \rangle,\sigma} \operatorname{iSWAP} (\theta = t \Delta t / n; \, \eta=0).
\end{align}

To efficiently implement the second part of the kinetic hopping term, we first need to swap the fermionic modes such that the pairs that we want to perform an operation on become neighbors on the qubit layout.
\begin{widetext}
\begin{align}
    e^{-i \Delta t H_3 / n}
    &= \exp \left[ -i \frac{\Delta t}{n} \cdot \left(-\frac{1}{2}t\right) \sum_{\langle i,j \rangle} \sum_{\sigma \in \{\uparrow,\downarrow\}} \left( X_{i,\sigma} X_{j,\sigma} + Y_{j,\sigma} Y_{i,\sigma} \right) \left( Z_{i+1,\sigma} Z_{i+2,\sigma} \dots Z_{j-1,\sigma} \right) \right] \nonumber \\
    &= \prod_{\langle i, j \rangle,\sigma} \exp \left[ -i \frac{(-t) \Delta t}{2n} \left( X_{i,\sigma} X_{j,\sigma} + Y_{i,\sigma} Y_{j,\sigma} \right) \left( Z_{i+1,\sigma} Z_{i+2,\sigma} \dots Z_{j-1,\sigma} \right) \right] \nonumber \\
    &= \prod_{\langle i^{\prime}, j^{\prime} \rangle,\sigma} \operatorname{fSWAP}_{i, i^{\prime}} \cdot \operatorname{fSWAP}_{j, j^{\prime}} \cdot \exp \left[ -i \frac{(-t) \Delta t}{2n} \left( X_{i^{\prime},\sigma} X_{j^{\prime},\sigma} + Y_{i^{\prime},\sigma} Y_{j^{\prime},\sigma} \right) \right] \cdot \operatorname{fSWAP}_{i, i^{\prime}} \cdot \operatorname{fSWAP}_{j, j^{\prime}} \nonumber \\
     &= \prod_{\langle i^{\prime}, j^{\prime} \rangle,\sigma} \operatorname{fSWAP}_{i, i^{\prime}} \cdot \operatorname{fSWAP}_{j, j^{\prime}} \cdot \operatorname{iSWAP}_{i^{\prime}, j^{\prime}} (\theta = 2t \Delta t / n; \, \eta=0) \cdot \operatorname{fSWAP}_{i, i^{\prime}} \cdot \operatorname{fSWAP}_{j, j^{\prime}}.
\end{align}
\end{widetext}

Here, $\langle i, j \rangle$ are all the colored neighboring pairs of qubits that are shown in configuration \protect\circled{2} of \cref{fig:time_evolution_op}, while $\langle i^{\prime}, j^{\prime} \rangle$ is the set of all neighboring qubit pairs as depicted in configuration \protect\circled{4}.
To implement the hopping operation between each of these pairs, no additional $Z$-strings are required since every pair of adjacent sites $i^{\prime}$ and $j^{\prime}$ in the physical model also refers to neighboring qubits in the Jordan-Wigner encoded model (see \cref{fig:time_evolution_op}). This is achieved by moving fermionic modes that are not adjacent into neighboring positions on the qubit grid. The fermionic modes are exchanged using the \gls{FSWAP} operator, 
\begin{equation}
\begin{quantikz}
    \ghost{X} & \swap{1} & \wire[l][1]["^{\sim}"{above,pos=1.0}]{a} \\
    \qw & \targX{} & \ghost{X} \qw
\end{quantikz}
=
\begin{pmatrix}
    1 & 0 & 0 & 0 \\
    0 & 0 & 1 & 0 \\
    0 & 1 & 0 & 0 \\
    0 & 0 & 0 & -1
\end{pmatrix},
\end{equation}
which is similar to a $\operatorname{SWAP}$ gate but accounts for the minus sign that arises when two fermionic modes are exchanged.

In \cref{sec:fswap_hopping} we explain how this approach of swapping the fermionic modes to neighboring sites and subsequently applying local hopping gates implicitly preserves the properties of the system and leads to the original Hamiltonian that was already introduced in the previous sections.

\subsection{Controlling the time evolution operator}
\label{sec:pauli_control}
So far, we have described the implementation of the forward time evolution operator (\cref{eq:time_evolution_op}). In this section, we provide the details of controlling the direction of the time evolution to implement the operator in \cref{eq:forward_backward_time_evolution_op}.

Since there is no unitary operator $K$ that anti-commutes with the entire Fermi-Hubbard Hamiltonian $H$, we must find a unitary operator $K_i$ for each term $H_i$ in the Hamiltonian. To reverse the direction of the time evolution, one Trotter step then becomes
\begin{align}
    S_2(\Delta t) &= K_1 \, e^{-i \frac{\Delta t}{2n} H_{1}} \, K_1 \cdot K_2 \, e^{-i \frac{\Delta t}{2n} H_{2}} \, K_2 \nonumber \\
    &\quad \cdot K_3 \, e^{-i \frac{\Delta t}{n} H_{3}} \, K_3 \nonumber \\
    &\quad \cdot K_2 \, e^{-i \frac{\Delta t}{2n} H_{2}} \, K_2 \cdot K_1 \, e^{-i \frac{\Delta t}{2n} H_{1}} \, K_1.
\end{align}
For the terms $H_2$ and $H_3$, we can find a unitary operator $K_2$ that anti-commutes with both terms. Hence, they cancel out, and the Trotter step becomes
\begin{align}
    S_2(\Delta t) &= K_1 \, e^{-i \frac{\Delta t}{2n} H_{1}} \, K_1 \nonumber \\
    &\quad \cdot K_2 \, e^{-i \frac{\Delta t}{2n} H_{2}} \, e^{-i \frac{\Delta t}{n} H_{3}} \, e^{-i \frac{\Delta t}{2n} H_{2}} \, K_2 \nonumber \\
    &\quad \cdot K_1 \, e^{-i \frac{\Delta t}{2n} H_{1}} \, K_1.
\end{align}
The unitary operators are
\begin{subequations}
\begin{equation}
    K_1 = X_{2,\uparrow} \otimes X_{3,\uparrow} \otimes X_{1,\downarrow} \otimes X_{4,\downarrow},
\end{equation}
and
\begin{equation}
    K_2 = Z_{2,\uparrow} \otimes Z_{3,\uparrow} \otimes Z_{1,\downarrow} \otimes Z_{4,\downarrow}.
\end{equation}
\end{subequations}
As we can see, $K_1$ and $K_2$ are just Pauli strings that can easily be implemented on the target device.

To control the direction of the time evolution, we need to control the unitary operators. Since the Pauli strings are both zero-controlled by the ancilla qubit, the final gates are zero-controlled X-gates and zero-controlled Z-gates, respectively. Those can be implemented on the target hardware using the circuit identities in \cref{sec:gate_decomposition}.

\subsection{Shifting the Hamiltonian}
\label{sec:hamiltonian_shift}
As already mentioned above, we need to shift the eigenvalues of the Hamiltonian to be within the interval $\left[\eta, \pi-\eta \right]$. This is achieved by performing an affine transformation on the Hamiltonian
\begin{equation}
    H_{\text{sh}} = c_1 H + c_2 I,
\end{equation}
where
\begin{equation}
    c_1 = \frac{\pi - 2\eta}{\lambda_{\text{max}} - \lambda_{\text{min}}}, \qquad c_2 = \eta - c_1 \lambda_{\text{min}},
\end{equation}
with $\lambda_{\text{min}}$ denoting the smallest eigenvalue corresponding to the ground state, and $\lambda_{\text{max}}$ being the highest valued eigenvalue corresponding to the highest excited energy state. The time evolution operator then becomes
\begin{equation}
\begin{split}
    U_{\text{sh}}& = \exp \left( -i \Delta t H_{\text{sh}} \right) \\
    &= \exp \left( -i \Delta t (c_1 H + c_2 I ) \right) \\
    &= \exp(-i c_2 \Delta t I) \exp (-i c_1 \Delta t H).
\end{split}
\end{equation}
Therefore, as first used in \cite{QETU2022}, the controlled forward and backward operator becomes
\begin{equation}
\begin{split}
    V_{\text{sh}} &=
    \begin{pmatrix}
        U_{\text{sh}}^{\dagger} & 0 \\
        0 & U_{\text{sh}}
    \end{pmatrix} \\
    &=
    \begin{pmatrix}
        e^{i c_2 \Delta t I} & 0 \\
        0 & e^{-i c_2 \Delta t I}
    \end{pmatrix}
    \begin{pmatrix}
        e^{i c_1 \Delta t H} & 0 \\
        0 & e^{-i c_1 \Delta t H}
    \end{pmatrix} \\
    &=
    \begin{quantikz}
        \qw & \gate{R_z(-2 c_2)} & \gate[2]{V(c_1 \Delta t)} & \qw \\
        \qw & & \qw & \qw
    \end{quantikz},
\end{split}
\end{equation}
which can be implemented physically by scaling the simulation time with $c_1$ and using a $ R_z$ rotation on the ancilla qubit to add a global phase in the respective time direction to the forward and backward time evolution operator.

\section{\label{sec:measurements}Measurements}
After preparing the ground state vector $\ket{\psi_0}$ with sufficiently high fidelity, we can extract the ground-state energy $E_0$ by conducting multiple measurements to estimate the expectation value of the ground state. The naïve method would be to repeat that energy measurement for every term $H_i$ in the Hamiltonian $H$, but it turns out that this process can be optimized by grouping commuting terms and measuring them in parallel. In the case of the Fermi-Hubbard model, it is possible to use the same grouping as in \cref{eq:fh_jw}.

The entire on-site term $H_1$ can be estimated by measuring the relevant qubit pairs (representing spin up and spin down) in the computational basis. In the Jordan-Wigner representation, a single on-site term becomes $Z \otimes Z$ which is already a diagonal matrix with eigenvalue $-1$ corresponding to eigenvectors $\ket{01}$ and $\ket{10}$, and eigenvalue $1$ corresponding to $\ket{00}$ and $\ket{11}$. Hence, the energy state of an on-site term at a particular site is equivalent to the probability of obtaining the same measurement result for both corresponding qubits minus the probability of getting two different outcomes.

The individual hopping terms are mapped to $\frac{1}{2} \left( XX + YY \right)$ during the Jordan-Wigner transformation. Since measurements are only possible with respect to the computational basis states, it is necessary to perform a change of basis prior the the measurements. As described in \cite{Cade2020}, these Jordan-Wigner-transformed hopping terms can be diagonalized like $\frac{1}{2} \left( XX + YY \right) = U^{\dagger} \left( \ket{01} \bra{01} - \ket{10} \bra{10} \right) U$ where
\begin{equation}
    U = 
    \begin{pmatrix}
        1 & 0 & 0 & 0 \\
        0 & \frac{1}{\sqrt{2}} & \frac{1}{\sqrt{2}} & 0 \\
        0 & \frac{1}{\sqrt{2}} & -\frac{1}{\sqrt{2}} & 0 \\
        0 & 0 & 0 & 1
    \end{pmatrix}
    =
    \begin{quantikz}
        \qw & \ctrl{1} & \gate{H} & \ctrl{1} & \qw\\
        \qw & \targ{} & \ctrl{-1} & \targ{} & \qw
    \end{quantikz}.
    \label{eq:hopping_basis_transform}
\end{equation}
Therefore, the expectation value of a hopping term acting on a qubit pair can be determined by subtracting the probability of measuring $10_2$ from the probability of obtaining $01_2$ as the measurement result. We have
\begin{align*}
    & \left\langle \psi_0 \middle| \frac{1}{2} \left( X_{i,\sigma} X_{j,\sigma} + Y_{i,\sigma} Y_{j,\sigma} \right) \middle| \psi_0 \right\rangle \\
    & \qquad = \left\langle \psi_0 \middle|  U^{\dagger} \left( \ket{01} \bra{01} - \ket{10} \bra{10} \right) U \middle| \psi_0 \right\rangle \\
    & \qquad = \left\langle \psi_0^{\prime} \middle|  \left( \ket{01} \bra{01} - \ket{10} \bra{10} \right)  \middle| \psi^{\prime}_0 \right\rangle \\
    & \qquad = \braket{\psi_0^{\prime}}{01} \braket{01}{\psi_0^{\prime}} - \braket{\psi_0^{\prime}}{10} \braket{10}{\psi_0^{\prime}} \\
    & \qquad = \left[ \mathds{P}(01) - \mathds{P}(10) \right],
\end{align*}
where $\mathds{P}(01)$ is the probability of measuring $01_2$, $\ket{\psi_0^{\prime}} = U \ket{\psi_0}$, and we have used the fact that $\mathds{P}(i) = \lvert \braket{i}{\psi^{\prime}_0} \rvert^2$.
Consequently, the first subset of hopping terms $H_2$ can be measured by first applying the transformation in \cref{eq:hopping_basis_transform} to all qubit pairs in configuration \protect\circled{6} of \cref{fig:time_evolution_op} and then performing the measurements and post-processing as described in the next section.
The second set of hopping terms $H_3$ can be determined in a similar way but with the additional need to swap the spin orbitals to be as in configuration \protect\circled{4} before the basis transformation and measurements. The reason is that otherwise, the terms of $H_2$ would anti-commute with $H_3$ and demolish the measurements with respect to the $(XX+YY)$-basis. Swapping the spin states has the supplementary effect of superseding the need to handle the additional $Z$-strings.

Therefore, the overall ground-state energy can be reconstructed by repeating the measurements only on three circuits:
\begin{widetext}
\begin{align*}
    E_0 &= \left\langle \psi_0 \middle| H \middle| \psi_0 \right\rangle \\
    &= \frac{1}{4}u \sum_{j=1}^{n} \left\langle \psi_0 \middle| Z_{j,\uparrow} Z_{j,\downarrow} \middle| \psi_0 \right\rangle
    - \frac{1}{2}t \sum_{\langle i,j \rangle} \sum_{\sigma \in \{ \uparrow,\downarrow \} } \left\langle \psi_0 \middle| \left( X_{i,\sigma} X_{j,\sigma} + Y_{i,\sigma} Y_{j,\sigma} \right) \middle| \psi_0 \right\rangle \\
    &= \frac{1}{4}u \sum_{j=1}^{n} \sum_{\sigma_1=0}^{1} \sum_{\sigma_{2}=0}^{1} (-1)^{\sigma_{1} + \sigma_{2}} \, \mathds{P}(q_{j,\uparrow}{=}\sigma_{1}, \, q_{j,\downarrow}{=}\sigma_{2} \mid Z_{j,\uparrow} Z_{j,\downarrow}) \\
    & \qquad - \frac{1}{2}t \sum_{\langle i,j \rangle} \sum_{\sigma \in \{ \uparrow,\downarrow \} } \mathds{P}(q_{i,\sigma}{=}0, \, q_{j,\sigma}{=}1 \mid U_{i,\sigma} U_{j,\sigma}) - \mathds{P}(q_{i,\sigma}{=}1, \, q_{j,\sigma}{=}0 \mid U_{i,\sigma} U_{j,\sigma}) 
\end{align*}
\end{widetext}
Here, $\mathds{P}(q_{1}{=}\sigma_{1}, \, q_{2}{=}\sigma_{2} \mid Z_{1} Z_{2})$ denotes the probability of measuring the qubit at position $q_1$ as $\ket{\sigma_1}$ and the qubit at position $q_2$ as $\ket{\sigma_2}$ in the computational basis (Z-basis). Likewise, $\mathds{P}(q_{1}{=}\sigma_{1}, \, q_{2}{=}\sigma_{2} \mid U_{1} U_{2})$ is the probability of obtaining these measurements after applying the basis transformation \cref{eq:hopping_basis_transform} on the qubit pair $q_1$, $q_2$.

\section{Error mitigation}
Quantum computers are highly susceptible to errors, and even in the fault-tolerant regime, noise and error sources are still present.
Therefore, it is crucial to implement techniques that can detect, reduce, or correct these errors to ensure accurate and reliable results.
Here, we focus on error mitigation schemes that exploit known properties and symmetries of the model.

We use a post-selection strategy to filter out or at least statistically reduce errors in the final result without modifying the physical qubits directly. This strategy leverages the physical symmetries and conservation properties of the Fermi-Hubbard Hamiltonian to discard measurement results that violate these laws, thereby reducing the impact of errors on the final outcome. In our case, the total number of fermions is conserved throughout the computation if the simulation starts with a specific number. Also, the fact that the auxiliary qubit must be in the zero state can be seen as an additional constraint to post-select the measurements.

\section{Numerical Experiments}
In this section, we present the numerical evaluation of the \gls{QETU} algorithm. The evaluation aims to analyze the algorithm's susceptibility to noise and error influences.
We use IBM's Qiskit \cite{Qiskit} device backend noise simulator for the experiments. We performed the simulations both on an ideal model and on various noisy models of the target hardware. The source code of our implementation is available at GitHub \cite{GitHub}.

In all scenarios, the Coulomb repulsion energy was set to $u=1$, and the kinetic hopping energy was set to $t=1$. For each forward and backward time evolution operator $V$ that was used in the \gls{QETU} circuit, only a single second-order Trotter step was used, which was sufficient to achieve a close approximation with an absolute error of only 0.00518.

\begin{figure}
    \centering
    \resizebox{\columnwidth}{!}{\begin{tikzpicture}[remember picture]
    \begin{axis}[
        xlabel=degree $d$,
		ylabel={overlap $\lvert \braket{\psi_{\text{final}}}{\psi_0} \rvert ^2$},
        legend pos=north west
    ]
    \addplot[color=red,mark=x] coordinates {
		(1, 0.12754541844314995)
        (2, 0.2251430871318981)
        (3, 0.19830686797950745)
        (4, 0.18361959032559536)
        (5, 0.2726705793213184)
        (6, 0.4424540382879922)
        (7, 0.3461443750921958)
        (8, 0.27695163921187904)
        (9, 0.3642376776090261)
        (10, 0.4908894305951951)
        (11, 0.4501480819542708)
        (12, 0.4422594877167402)
        (13, 0.5045280366798626)
        (14, 0.5903297929928335)
        (15, 0.6639109609275878)
        (16, 0.7053128820871332)
        (17, 0.7641934264951178)
        (18, 0.8261507812695079)
        (19, 0.8527778354934316)
        (20, 0.8843590465463045)
        (21, 0.9116497231989356)
        (22, 0.9330799636068442)
        (23, 0.9536583134307263)
        (24, 0.9559231119569428)
        (25, 0.9500897316827167)
        (26, 0.9517575390145597)
        (27, 0.9584334950857594)
        (28, 0.9685348911633774)
        (29, 0.9765134050707543)
        (30, 0.9825104700770186)
        (31, 0.9873112827381483)
        (32, 0.9887776869662911)
        (33, 0.9885591819055927)
        (34, 0.9897722969164732)
        (35, 0.9904986312272467)
        (36, 0.9905113926910657)
        (37, 0.9907043707377536)
        (38, 0.9910023179132037)
        (39, 0.9914672122691003)
        (40, 0.9918636959292754)
        (41, 0.9923809369584657)
        (42, 0.9925163216855097)
        (43, 0.9907741864539811)
        (44, 0.9892542236803126)
        (45, 0.9898175590587025)
        (46, 0.9906088402438251)
        (47, 0.9906590017038192)
        (48, 0.9908141474010533)
        (49, 0.9916241412137841)
        (50, 0.9923340989237746)
	};
    \addlegendentry{noiseless}
    \coordinate (insetPosition) at (rel axis cs:0.95,0.05);
    \coordinate (c1) at (axis cs:45,0.98);
    \coordinate (c2) at (axis cs:50,1);
    \draw (c1) rectangle (axis cs:50,1);
    \end{axis}
    
    \begin{axis}[
        at={(insetPosition)},
        anchor={outer south east},
        footnotesize,
        name=ax2,
        yticklabel style={/pgf/number format/.cd,fixed,precision=3,zerofill},
        width=0.25*\textwidth,
    ]
    \addplot[color=red,mark=x] coordinates {
        (45, 0.9898175590587025)
        (46, 0.9906088402438251)
        (47, 0.9906590017038192)
        (48, 0.9908141474010533)
        (49, 0.9916241412137841)
        (50, 0.9923340989237746)
	};
    \end{axis}

    \draw [dashed] (c2) -- (ax2.north east);
    \draw [dashed] (c1) -- (ax2.north west);
\end{tikzpicture}}
    \caption{Results of the numerical simulations in the ideal, noiseless case. The graph shows the overlap of the quantum state $\ket{\psi_{\text{final}}}$ that is prepared by the presented circuit, with the expected ground state $\ket{\psi_{0}}$ using a polynomial approximation of the step function with increasing degree $d$. Only one single Trotter step is used.}
    \label{fig:ground_state_overlap}
\end{figure}
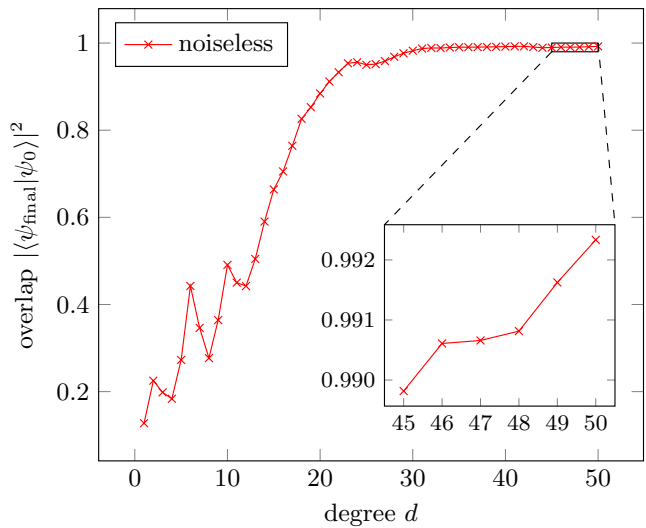

Firstly, we are interested in how well the algorithm can prepare the ground state of the Hamiltonian. Therefore, we simulated the statevector evolution and calculated the overlap with the expected ground state as $\lvert \braket{\psi_{\text{final}}}{\psi_0} \rvert ^2$ where $\ket{\psi_{\text{final}}}$ denotes the final state that was prepared by the \gls{QETU} algorithm. The reference ground state vector $\ket{\psi_{0}}$ was obtained through exact diagonalization of the Hamiltonian. In every setup, the circuit was initialized with the initial state $\ket{0} \ket{\psi_{\text{init}}}$. We set the initial state to $\ket{\psi_{\text{init}}} = \ket{1001}\ket{\hbox{\texttt{-}}\hbox{\texttt{-}}\hbox{\texttt{+}}\hbox{\texttt{+}}}$, which was determined empirically to guarantee a sufficiently large initial overlap $\gamma = \vert \braket{\psi_0}{\psi_{\text{init}}} \lvert \ge 0.09102$. The results of the numerical simulation for the ground state preparation are presented in \cref{fig:ground_state_overlap}.

\begin{figure*}
    \centering
    \resizebox{2\columnwidth}{!}{\input{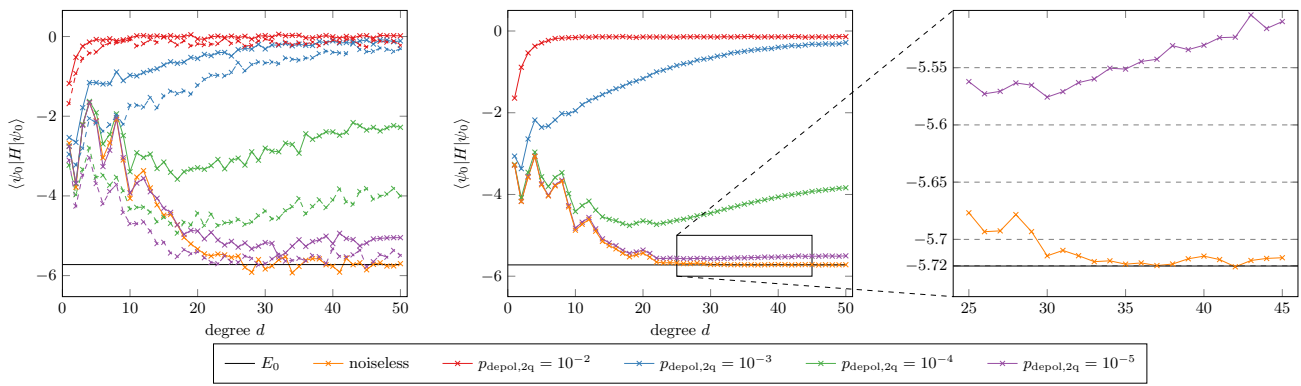}}
    \caption{Estimation of the ground-state energy of the Fermi-Hubbard model with $u=1$ and $t=1$ using the QETU algorithm with increasing polynomial degree $d$ and a single Trotter step. Plotted are the results for different noise models with and without error mitigation. (Left) Each experiment is repeated 10.000 times and (Middle) $10^7$ times. Here, $p_{\text{depol,2q}}$ denotes the depolarizing probability for two-qubit gates; the depolarizing probability for single-qubit gates was set to $p_{\text{depol,1q}} = p_{\text{depol,2q}} / 10$ and the measurement error probability was set to $p_{\text{depol,meas}} = 10 \cdot p_{\text{depol,2q}}$. As visible in the magnification of the bottom right part of the middle figure, the approximation error remains around $10^{-2}$ even in the ideal case due to the stochastic limit of the direct expectation value measurement method.}
    \label{fig:energy_depol}
\end{figure*}

Secondly, we investigated the performance of the algorithm in estimating the ground-state energy. As described in \cref{sec:measurements}, the energy estimation process requires repeating the measurements for three circuits to account for the different bases of the energy components. Both circuits were initialized with the same initial state as for the statevector simulation. For each circuit, the measurements were then repeated 10.000 times to study the performance of the algorithm with a realistic amount of experiments, as well as $10^7$ times to obtain the statistical limit. The results are visualized in \cref{fig:energy_depol}.

The noisy simulations were executed using the Qiskit AerSimulator.
Our noise model accounts for depolarizing noise and measurement errors.
Depolarizing noise is a type of quantum noise in which a qubit gets replaced with the maximally mixed state with some probability $p$ due to infidelities of the quantum gates \cite{Georgopoulos2021, Nielsen_Chuang_2010}. We conducted multiple simulations with varying depolarization probabilities. The two-qubit error rate was set to $p_{\text{depol,2q}}$, while the single-qubit error rate was set to $p_{\text{depol,1q}} = p_{\text{depol,2q}} / 10$ and the measurement error rate to $p_{\text{depol,meas}} = 10 \cdot p_{\text{depol,2q}}$.

\section{Discussion and Outlook}

In this work, we showed that by utilizing the technique of fermionic swap networks, we can efficiently realize a controlled forward and backward time evolution operator of the Fermi-Hubbard model on a 2D lattice. This time evolution operator serves as a key ingredient to preparing the ground state of that model with the \gls{QETU} algorithm. We considered the $2 \times 2$ Fermi-Hubbard model as a toy model to study what hardware resources are needed for a practical implementation.

In the case of the $2 \times 2$ Fermi-Hubbard model, the success of the ground state preparation is mostly dependent on the degree of the polynomial approximation of the step function, as can be seen in \cref{fig:ground_state_overlap}. For the toy model, a single Trotter step yields an approximation that is already accurate enough to deliver $\ge 0.99$ fidelity of the prepared state under the condition that the polynomial degree is high enough (see \cref{fig:trotter_error}). The reason is that the simulation time gets scaled down by the $c_1$ constant that is used to shift the spectrum of the Hamiltonian, and a smaller time step naturally corresponds to a more accurate Trotter approximation. In principle, the simulation time can be arbitrarily small; however, scaling down the time step also shrinks the range of the spectrum of the Hamiltonian and, thus, decreases the spectral gap. This necessitates a sharper transition of the step function, which can only be achieved through a higher polynomial degree. Because the spectral gap is expected to decrease with increasing system sizes, simulating those systems not only requires more qubits but also comes with an increased circuit depth. Additionally, finding the best trade-off between lower-depth Trotter approximations and better polynomial approximations becomes critical to the success of the \gls{QETU} algorithm in these situations. The algorithm's scalability can be improved by successively scaling the Hamiltonian to increase the range of the spectrum and applying a fixed degree polynomial to filter out all unwanted states as proposed by \cite{Karacan2024}.

As shown in the left plot of \cref{fig:energy_depol}, the gate fidelities required to estimate the ground state energy of our toy model with reasonable accuracy exceed the current capabilities of today’s superconducting quantum computers. Since decoherence times generally pose a more significant challenge than qubit count, a promising direction for future work is to reduce the circuit depth by parallelizing the \gls{QETU} algorithm through polynomial factorization, as recently proposed in \cite{Martyn2024}.

Furthermore, the attainable accuracy of the estimation is bounded by the stochastic limit of the direct expectation value measurement. \cref{fig:energy_depol} shows that a realistic, limited number of shots cannot achieve a sensible ground state energy estimation accuracy even in the ideal noiseless case. In future work, more sophisticated techniques could be used to achieve higher precision with a limited amount of shots.

Moreover, the success of the \gls{QETU} algorithm depends on a high enough overlap between the initial state and the ground state. Our experiments suggest that random initialization of the initial state cannot guarantee this overlap, often leading to a low probability of convergence to the ground state. We carefully selected the initial state in the toy model presented in this work. Other ways to prepare the initial state are indispensable for practical applications with larger systems.

In this work, we focused on implementing and optimizing the \gls{QETU} circuit for a specific instance of the Fermi-Hubbard model. Not all of the optimizations presented here generalize to larger system sizes. In particular, the small system size of the toy model allows for the placement of the auxiliary qubit in the center of the qubit grid. In that position, the auxiliary qubit is adjacent to every fermionic mode in one of only two distinct fermion-to-qubit mappings. As a result, the controlled Pauli operators can be implemented on the same mapping as the associated Hamiltonian terms. This is generally not the case when implementing larger systems. Here, finding the optimal Pauli strings is non-trivial as it is interconnected with finding a decomposition of the Hamiltonian into smaller groups of weighted Pauli operators such that the auxiliary qubit can interact locally with the qubits associated with the respective terms while minimizing the number of required fermion-to-qubit mappings.

Lastly, it is important to remember that in this work, we assume knowledge of the parameter $\mu$ obtained by direct diagonalization of the Hamiltonian. To estimate the ground state energy without a priori knowledge of $\mu$, we need to perform a binary search on the Hamiltonian spectrum and repeatedly apply the \gls{QETU} algorithm described in the original paper \cite{Dong2022}.

\begin{acknowledgments}
We acknowledge the support and funding received for the MUNIQC-SC initiative under funding number 13N16188 from the VDI Technology Center as part of the German BMBF program.
\end{acknowledgments}

\clearpage
\onecolumngrid

\appendix

\section{The Jordan-Wigner transformation of the 2D Fermi-Hubbard model}
\label{sec:jw_transform}
The Jordan-Wigner transformation is a method to map fermionic systems onto qubit systems that can be run on a quantum computer.
The fermionic operators are transformed like
\begin{equation}
    a_i \mapsto \left( \bigotimes_{k=1}^{i-1} Z_k \right) \Sigma^{+}_i,
    \qquad
    a^{\dagger}_j \mapsto \left( \bigotimes_{k=1}^{j-1} Z_k \right) \Sigma^{-}_j,
\label{eq:jw}
\end{equation}
\\
with $\Sigma^{+}_j = \ket{0}\bra{1} = \frac{1}{2} (X_j + i Y_j)$ and $\Sigma^{-}_j = \ket{1}\bra{0} = \frac{1}{2} (X_j - i Y_j)$.
Therefore, the number operator is mapped via
\begin{align}
    n_{i,\sigma} &= a^{\dagger}_{i,\sigma} a_{i,\sigma} \\
    &\mapsto \left( \bigotimes_{k=1}^{i-1} Z_{k,\sigma} \right) \Sigma^{-}_{i,\sigma} \left( \bigotimes_{k=1}^{i-1} Z_{k,\sigma} \right) \Sigma^{+}_{i,\sigma} \\
    &= \Sigma^{-}_{i,\sigma} \Sigma^{+}_{i,\sigma} = \frac{1}{2} \left( I_{i,\sigma} - Z_{i,\sigma} \right).
\end{align}
\\
Hence, under the Jordan-Wigner transformation, the onsite term becomes
\begin{align}
    u \sum_{i=1}^n n_{i, \uparrow} n_{i, \downarrow}
    &\mapsto \frac{1}{4} u \sum_{i=1}^n \left( I_{i,\uparrow} - Z_{i,\uparrow} \right) \cdot \left( I_{i,\downarrow} - Z_{i,\downarrow} \right).
\end{align}
\\
Often, additional terms are added in order to simplify the calculations
\begin{equation}
     u \sum_{i=1}^n \left( n_{i, \uparrow} - \frac{1}{2} \right) \cdot \left( n_{i, \downarrow} - \frac{1}{2} \right)
    \mapsto \frac{1}{4} u \sum_{i=1}^n Z_{i,\uparrow} Z_{i,\downarrow}.
\end{equation}
\\
Without loss of generality, we assume that $i \le j$. Using \cref{eq:jw}, we arrive at
\begin{equation}
    a^{\dagger}_i a_j \mapsto (\Sigma^{-}_i Z_i) \left( \bigotimes_{k=i+1}^{j-1} Z_k \right) \Sigma^{+}_j.
\end{equation}
\\
Each hopping operator is mapped via
\begin{align}
    h_{i,j,\sigma} &= a^{\dagger}_{i,\sigma} a_{j,\sigma} + a^{\dagger}_{j,\sigma} a_{i,\sigma} \\
    &\mapsto \left( \Sigma^{-}_{i,\sigma} Z_{i,\sigma} \right) \left( \bigotimes_{k=i+1}^{j-1} Z_{k,\sigma} \right) \Sigma^{+}_{j,\sigma} + \left( Z_{i,\sigma} \Sigma^{+}_{i,\sigma} \right) \left( \bigotimes_{k=i+1}^{j-1} Z_{k,\sigma} \right) \Sigma^{-}_{j,\sigma} \\
    &= \Sigma^{-}_{i,\sigma} \left( \bigotimes_{k=i+1}^{j-1} Z_{k,\sigma} \right) \Sigma^{+}_{j,\sigma} + \Sigma^{+}_{i,\sigma} \left( \bigotimes_{k=i+1}^{j-1} Z_{k,\sigma} \right) \Sigma^{-}_{j,\sigma} \\
    &= \frac{1}{2} \left[ X_{i,\sigma} \left( \bigotimes_{k=i+1}^{j-1} Z_{k,\sigma} \right) X_{j,\sigma} + Y_{i,\sigma} \left( \bigotimes_{k=i+1}^{j-1} Z_{k,\sigma} \right) Y_{j,\sigma} \right]
\end{align}
and so we arrive at the overall hopping term as
\begin{align}
    -t \sum_{\langle j,l \rangle} \sum_{\sigma \in \{\uparrow,\downarrow\}} h_{i,j,\sigma}
    \mapsto
    - \frac{1}{2} t \sum_{\langle j,l \rangle} \sum_{\sigma \in \{\uparrow,\downarrow\}}
    \left( X_{i,\sigma}  X_{j,\sigma} + Y_{i,\sigma} Y_{j,\sigma} \right) \left( Z_{i+1,\sigma} Z_{i+2,\sigma} \dots Z_{j-1,\sigma} \right).
\end{align}

\section{Implementation of long-range hopping terms through fSWAP gates}
\label{sec:fswap_hopping}
In this section, we want to build an intuition as to why our method of swapping distant fermionic modes to neighboring sites has the same effect as introducing additional $Z$-strings.

As an example, consider a hopping term $\operatorname{hop}_{1,3}$ between two non-local qubits $q_1$ and $q_2$. The term $\operatorname{hop}_{1,3}$ can also be expressed by switching the fermionic modes of $q_1$ and $q_2$, applying the hopping term on the neighboring qubits $q_2$ and $q_3$, and subsequently switching the fermionic modes back to the original configuration: 
\begin{equation}
\begin{quantikz}[row sep={1cm,between origins}]
    \lstick{$q_1$} & \gate[style={fill=blue!20}]{XY}\wire[d][2]{q} & \qw \\
    \lstick{$q_2$} & & \qw \\
    \lstick{$q_3$} & \gate[style={fill=blue!20}]{XY} & \qw
    \end{quantikz}
=
\begin{quantikz}[row sep={1cm,between origins}]
    \ghost{XY} & \swap{1} & \wire[l][1]["^{\sim}"{above,pos=1.0}]{a} & \swap{1} & \wire[l][1]["^{\sim}"{above,pos=1.0}]{a} \\
    \qw & \targX{} & \gate[style={fill=blue!20}]{XY}\wire[d][1]{q}  & \targX{} & \qw \\
    \qw & \qw & \gate[style={fill=blue!20}]{XY} & \qw & \qw
\end{quantikz}
\label{eq:fswap_hopping}
\end{equation}
We show that \cref{eq:fswap_hopping} holds true by explicitly calculating the exponential form of the right-hand side of the equation. This calculation follows a similar approach as in the Appendix of \cite{Kökcü2023}.
The hopping operator between $q_2$ and $q_3$ is defined as
\begin{equation}
    \operatorname{hop}_{2,3} = \exp \left[ +i \frac{t \Delta t}{n} \left( X_2 X_3 + Y_2 Y_3 \right) \right].
\end{equation}
The fermionic swap operator is
\begin{equation}
    \operatorname{fSWAP} = e^{-i \frac{\pi}{4} (XX + YY)} e^{-i \frac{\pi}{2} ZZ} e^{+i \frac{\pi}{4} IZ} e^{+i \frac{\pi}{4} ZI},
\end{equation}
and it is easy to verify that $\operatorname{fSWAP} = \operatorname{fSWAP}^{\dagger}$.
Plugging it all in, we get:
\begin{align}
    &\operatorname{fSWAP}_{1,2} \cdot \operatorname{hop}_{2,3} \cdot \operatorname{fSWAP}^{\dagger}_{1,2} \nonumber \\
    &= e^{-i \frac{\pi}{4} (X_1 X_2 + Y_1 Y_2)} e^{-i \frac{\pi}{2} Z_1 Z_2} e^{+i \frac{\pi}{4} I_1 Z_2} e^{+i \frac{\pi}{4} Z_1 I_2} \cdot e^{+i \frac{t \Delta t}{n} ( X_2 X_3 + Y_2 Y_3)} \cdot  e^{+i \frac{\pi}{4} (X_1 X_2 + Y_1 Y_2)} e^{+i \frac{\pi}{2} Z_1 Z_2} e^{-i \frac{\pi}{4} I_1 Z_2} e^{-i \frac{\pi}{4} Z_1 I_2} \nonumber \\
    &= e^{-i \frac{\pi}{4} X_1 X_2} e^{-i \frac{\pi}{4} Y_1 Y_2} e^{-i \frac{\pi}{2} Z_1 Z_2} e^{+i \frac{\pi}{4} I_1 Z_2} e^{+i \frac{\pi}{4} Z_1 I_2} \cdot e^{+i \frac{t \Delta t}{n} ( X_2 X_3 + Y_2 Y_3)} \cdot e^{-i \frac{\pi}{4} Z_1 I_2} e^{-i \frac{\pi}{4} I_1 Z_2} e^{+i \frac{\pi}{2} Z_1 Z_2} e^{+i \frac{\pi}{4} Y_1 Y_2} e^{+i \frac{\pi}{4} X_1 X_2} \nonumber \\
    &= e^{-i \frac{\pi}{4} X_1 X_2} e^{-i \frac{\pi}{4} Y_1 Y_2} e^{-i \frac{\pi}{2} Z_1 Z_2} e^{+i \frac{\pi}{4} I_1 Z_2} \cdot e^{+i \frac{t \Delta t}{n} ( X_2 X_3 + Y_2 Y_3)} \cdot e^{-i \frac{\pi}{4} I_1 Z_2} e^{+i \frac{\pi}{2} Z_1 Z_2} e^{+i \frac{\pi}{4} Y_1 Y_2} e^{+i \frac{\pi}{4} X_1 X_2} \nonumber \\
    &= e^{-i \frac{\pi}{4} X_1 X_2} e^{-i \frac{\pi}{4} Y_1 Y_2} e^{-i \frac{\pi}{2} Z_1 Z_2} \cdot e^{+i \frac{t \Delta t}{n} ( -Y_2 X_3 + X_2 Y_3)} \cdot e^{+i \frac{\pi}{2} Z_1 Z_2} e^{+i \frac{\pi}{4} Y_1 Y_2} e^{+i \frac{\pi}{4} X_1 X_2} \nonumber \\
    &= e^{-i \frac{\pi}{4} X_1 X_2} e^{-i \frac{\pi}{4} Y_1 Y_2} \cdot e^{+i \frac{t \Delta t}{n} ( Y_2 X_3 - X_2 Y_3)} \cdot e^{+i \frac{\pi}{4} Y_1 Y_2} e^{+i \frac{\pi}{4} X_1 X_2} \nonumber \\
    &= e^{-i \frac{\pi}{4} X_1 X_2} \cdot e^{+i \frac{t \Delta t}{n} ( I_1 Y_2 X_3 - Y_1 Z_2 Y_3)} \cdot e^{+i \frac{\pi}{4} X_1 X_2} \nonumber \\
    &= e^{+i \frac{t \Delta t}{n} ( X_1 Z_2 X_3 - Y_1 Z_2 Y_3)} \nonumber \\
    &= \operatorname{hop}_{1,3} \cdot Z_2
\end{align}

\section{Gate decompositions}
\label{sec:gate_decomposition}
Here, we show decompositions of gates into our native gate set. Equivalences that only hold up to an irrelevant global phase factor are denoted with "$\approx$". The decompositions have been found using a depth-optimal brute-force search.
\begin{equation}
\begin{quantikz}
     \ghost{\frac{1}{2}} & \swap{1} & \wire[l][1]["^{\sim}"{above,pos=1.0}]{a} \\
    \qw & \targX{} & \ghost{\frac{1}{2}} \qw
\end{quantikz}
\approx
\begin{quantikz}
    \qw & \push{\bigotimes} \wire[d][1]["(\pi ; \, 0)"]{q} & \gate{R_z(-\frac{\pi}{2})} & \qw \\
    \qw & \push{\bigotimes} & \gate{R_z(-\frac{\pi}{2})} & \qw
\end{quantikz}
\end{equation}

\begin{equation}
\begin{quantikz}
    \qw & \gate[style={fill=teal!40}]{Z}\wire[d][1]{q} & \ghost{R_z(\theta)} \qw \\
    \qw & \gate[style={fill=teal!40}]{Z} & \ghost{R_z(\theta)} \qw
\end{quantikz}
=
\begin{quantikz}
    \qw & \gate[2]{R_{zz}(\theta)} & \qw \\
    \qw & & \qw
\end{quantikz}
\approx
\begin{quantikz}
    \qw & \gate{R_z(\theta)} & \ctrl{1} & \qw \\
    \qw & \gate{R_z(\theta)} & \gate{P(-2\theta)} & \qw
\end{quantikz}
\end{equation}

\begin{equation}
\begin{quantikz}
    \qw & \octrl{1} & \qw \\
    \qw & \gate{Z} & \qw
\end{quantikz}
\approx
\begin{quantikz}
    \qw & \qw  & \ctrl{1} & \qw \\
    \qw & \gate{R_z(\pi)} & \gate{P(\pi)} & \qw
\end{quantikz}
\end{equation}

\begin{equation}
\begin{quantikz}
    \qw & \octrl{1} & \ghost{R_z(\theta)} \qw \\
    \qw & \targ{} & \ghost{R_z(\theta)} \qw
\end{quantikz}
\approx
\begin{quantikz}
    \qw & \gate{R_z(\pi)} & \ctrl{1} & \qw & \qw & \qw \\
    \qw & \gate{\sqrt{Y}} & \gate{P(-\pi)} & \gate{\sqrt{Y}} & \gate{R_z(\pi)} & \qw
\end{quantikz}
\end{equation}

Also note that in \cref{fig:time_evolution_op}, the hopping terms can be natively implemented using a parameterized $\operatorname{iSWAP}$ gate:
\begin{equation}
\begin{quantikz}
    \qw & \gate[style={fill=blue!20}]{XY}\wire[d][1]{q} & \qw \\
    \qw & \gate[style={fill=blue!20}]{XY} & \qw
\end{quantikz}
=
\begin{quantikz}
    \qw & \gate[style={fill=red!20}]{XY}\wire[d][1]{q} & \qw \\
    \qw & \gate[style={fill=red!20}]{XY} & \qw
\end{quantikz}
=
\begin{quantikz}
    \qw & \push{\bigotimes} \wire[d][1]["(\theta; \, 0)"]{q} & \ghost{XY} \qw \\
    \qw & \push{\bigotimes} & \ghost{XY} \qw
\end{quantikz}
\end{equation}
Here, either $\theta = t\Delta t / n$ or $\theta = 2t\Delta t / n$, depending on whether the gate is used to implement the time evolution of $H_2$ or $H_3$.

\twocolumngrid


\bibliography{apssamp}

\providecommand{\noopsort}[1]{}\providecommand{\singleletter}[1]{#1}%
\begin{thebibliography}{25}%
\makeatletter
\providecommand \@ifxundefined [1]{%
 \@ifx{#1\undefined}
}%
\providecommand \@ifnum [1]{%
 \ifnum #1\expandafter \@firstoftwo
 \else \expandafter \@secondoftwo
 \fi
}%
\providecommand \@ifx [1]{%
 \ifx #1\expandafter \@firstoftwo
 \else \expandafter \@secondoftwo
 \fi
}%
\providecommand \natexlab [1]{#1}%
\providecommand \enquote  [1]{``#1''}%
\providecommand \bibnamefont  [1]{#1}%
\providecommand \bibfnamefont [1]{#1}%
\providecommand \citenamefont [1]{#1}%
\providecommand \href@noop [0]{\@secondoftwo}%
\providecommand \href [0]{\begingroup \@sanitize@url \@href}%
\providecommand \@href[1]{\@@startlink{#1}\@@href}%
\providecommand \@@href[1]{\endgroup#1\@@endlink}%
\providecommand \@sanitize@url [0]{\catcode `\\12\catcode `\$12\catcode
  `\&12\catcode `\#12\catcode `\^12\catcode `\_12\catcode `\%12\relax}%
\providecommand \@@startlink[1]{}%
\providecommand \@@endlink[0]{}%
\providecommand \url  [0]{\begingroup\@sanitize@url \@url }%
\providecommand \@url [1]{\endgroup\@href {#1}{\urlprefix }}%
\providecommand \urlprefix  [0]{URL }%
\providecommand \Eprint [0]{\href }%
\providecommand \doibase [0]{https://doi.org/}%
\providecommand \selectlanguage [0]{\@gobble}%
\providecommand \bibinfo  [0]{\@secondoftwo}%
\providecommand \bibfield  [0]{\@secondoftwo}%
\providecommand \translation [1]{[#1]}%
\providecommand \BibitemOpen [0]{}%
\providecommand \bibitemStop [0]{}%
\providecommand \bibitemNoStop [0]{.\EOS\space}%
\providecommand \EOS [0]{\spacefactor3000\relax}%
\providecommand \BibitemShut  [1]{\csname bibitem#1\endcsname}%
\let\auto@bib@innerbib\@empty
\bibitem [{\citenamefont {Dong}\ \emph {et~al.}(2022)\citenamefont {Dong},
  \citenamefont {Lin},\ and\ \citenamefont {Tong}}]{Dong2022}%
  \BibitemOpen
  \bibfield  {author} {\bibinfo {author} {\bibfnamefont {Y.}~\bibnamefont
  {Dong}}, \bibinfo {author} {\bibfnamefont {L.}~\bibnamefont {Lin}},\ and\
  \bibinfo {author} {\bibfnamefont {Y.}~\bibnamefont {Tong}},\ }\bibfield
  {title} {\bibinfo {title} {Ground-state preparation and energy estimation on
  early fault-tolerant quantum computers via quantum eigenvalue transformation
  of unitary matrices},\ }\href {https://doi.org/10.1103/PRXQuantum.3.040305}
  {\bibfield  {journal} {\bibinfo  {journal} {PRX Quantum}\ }\textbf {\bibinfo
  {volume} {3}},\ \bibinfo {pages} {040305} (\bibinfo {year}
  {2022})}\BibitemShut {NoStop}%
\bibitem [{\citenamefont {Gily\'{e}n}\ \emph {et~al.}(2019)\citenamefont
  {Gily\'{e}n}, \citenamefont {Su}, \citenamefont {Low},\ and\ \citenamefont
  {Wiebe}}]{Gilyen2019}%
  \BibitemOpen
  \bibfield  {author} {\bibinfo {author} {\bibfnamefont {A.}~\bibnamefont
  {Gily\'{e}n}}, \bibinfo {author} {\bibfnamefont {Y.}~\bibnamefont {Su}},
  \bibinfo {author} {\bibfnamefont {G.~H.}\ \bibnamefont {Low}},\ and\ \bibinfo
  {author} {\bibfnamefont {N.}~\bibnamefont {Wiebe}},\ }\bibfield  {title}
  {\bibinfo {title} {Quantum singular value transformation and beyond:
  exponential improvements for quantum matrix arithmetics},\ }in\ \href
  {https://doi.org/10.1145/3313276.3316366} {\emph {\bibinfo {booktitle}
  {Proceedings of the 51st Annual ACM SIGACT Symposium on Theory of
  Computing}}},\ \bibinfo {series and number} {STOC 2019}\ (\bibinfo
  {publisher} {Association for Computing Machinery},\ \bibinfo {address} {New
  York, NY, USA},\ \bibinfo {year} {2019})\ p.\ \bibinfo {pages}
  {193–204}\BibitemShut {NoStop}%
\bibitem [{\citenamefont {Low}\ and\ \citenamefont {Chuang}(2019)}]{Low2019}%
  \BibitemOpen
  \bibfield  {author} {\bibinfo {author} {\bibfnamefont {G.~H.}\ \bibnamefont
  {Low}}\ and\ \bibinfo {author} {\bibfnamefont {I.~L.}\ \bibnamefont
  {Chuang}},\ }\bibfield  {title} {\bibinfo {title} {Hamiltonian {S}imulation
  by {Q}ubitization},\ }\href {https://doi.org/10.22331/q-2019-07-12-163}
  {\bibfield  {journal} {\bibinfo  {journal} {{Quantum}}\ }\textbf {\bibinfo
  {volume} {3}},\ \bibinfo {pages} {163} (\bibinfo {year} {2019})}\BibitemShut
  {NoStop}%
\bibitem [{\citenamefont {Martyn}\ \emph {et~al.}(2021)\citenamefont {Martyn},
  \citenamefont {Rossi}, \citenamefont {Tan},\ and\ \citenamefont
  {Chuang}}]{Martyn2021}%
  \BibitemOpen
  \bibfield  {author} {\bibinfo {author} {\bibfnamefont {J.~M.}\ \bibnamefont
  {Martyn}}, \bibinfo {author} {\bibfnamefont {Z.~M.}\ \bibnamefont {Rossi}},
  \bibinfo {author} {\bibfnamefont {A.~K.}\ \bibnamefont {Tan}},\ and\ \bibinfo
  {author} {\bibfnamefont {I.~L.}\ \bibnamefont {Chuang}},\ }\bibfield  {title}
  {\bibinfo {title} {Grand unification of quantum algorithms},\ }\href
  {https://doi.org/10.1103/PRXQuantum.2.040203} {\bibfield  {journal} {\bibinfo
   {journal} {PRX Quantum}\ }\textbf {\bibinfo {volume} {2}},\ \bibinfo {pages}
  {040203} (\bibinfo {year} {2021})}\BibitemShut {NoStop}%
\bibitem [{\citenamefont {Schuch}\ and\ \citenamefont
  {Siewert}(2003)}]{Schuch2003}%
  \BibitemOpen
  \bibfield  {author} {\bibinfo {author} {\bibfnamefont {N.}~\bibnamefont
  {Schuch}}\ and\ \bibinfo {author} {\bibfnamefont {J.}~\bibnamefont
  {Siewert}},\ }\bibfield  {title} {\bibinfo {title} {Natural two-qubit gate
  for quantum computation using the $\mathrm{XY}$ interaction},\ }\href
  {https://doi.org/10.1103/PhysRevA.67.032301} {\bibfield  {journal} {\bibinfo
  {journal} {Phys. Rev. A}\ }\textbf {\bibinfo {volume} {67}},\ \bibinfo
  {pages} {032301} (\bibinfo {year} {2003})}\BibitemShut {NoStop}%
\bibitem [{\citenamefont {Abrams}\ \emph {et~al.}(2020)\citenamefont {Abrams},
  \citenamefont {Didier}, \citenamefont {Johnson}, \citenamefont {da~Silva},\
  and\ \citenamefont {Ryan}}]{Abrams2020}%
  \BibitemOpen
  \bibfield  {author} {\bibinfo {author} {\bibfnamefont {D.~M.}\ \bibnamefont
  {Abrams}}, \bibinfo {author} {\bibfnamefont {N.}~\bibnamefont {Didier}},
  \bibinfo {author} {\bibfnamefont {B.~R.}\ \bibnamefont {Johnson}}, \bibinfo
  {author} {\bibfnamefont {M.~P.}\ \bibnamefont {da~Silva}},\ and\ \bibinfo
  {author} {\bibfnamefont {C.~A.}\ \bibnamefont {Ryan}},\ }\bibfield  {title}
  {\bibinfo {title} {Implementation of {XY} entangling gates with a single
  calibrated pulse},\ }\href
  {https://doi.org/https://doi.org/10.1038/s41928-020-00498-1} {\bibfield
  {journal} {\bibinfo  {journal} {Nat. Electron.}\ }\textbf {\bibinfo {volume}
  {3}},\ \bibinfo {pages} {744–750} (\bibinfo {year} {2020})}\BibitemShut
  {NoStop}%
\bibitem [{\citenamefont {Hubbard}(1963)}]{Hubbard1963}%
  \BibitemOpen
  \bibfield  {author} {\bibinfo {author} {\bibfnamefont {J.}~\bibnamefont
  {Hubbard}},\ }\bibfield  {title} {\bibinfo {title} {Electron correlations in
  narrow energy bands},\ }\href
  {https://doi.org/https://doi.org/10.1098/rspa.1963.0204} {\bibfield
  {journal} {\bibinfo  {journal} {Proc. R. Soc. Lond. A}\ }\textbf {\bibinfo
  {volume} {276}},\ \bibinfo {pages} {238–257} (\bibinfo {year}
  {1963})}\BibitemShut {NoStop}%
\bibitem [{\citenamefont {Kivlichan}\ \emph {et~al.}(2018)\citenamefont
  {Kivlichan}, \citenamefont {McClean}, \citenamefont {Wiebe}, \citenamefont
  {Gidney}, \citenamefont {Aspuru-Guzik}, \citenamefont {Chan},\ and\
  \citenamefont {Babbush}}]{Kivlichan2018}%
  \BibitemOpen
  \bibfield  {author} {\bibinfo {author} {\bibfnamefont {I.~D.}\ \bibnamefont
  {Kivlichan}}, \bibinfo {author} {\bibfnamefont {J.}~\bibnamefont {McClean}},
  \bibinfo {author} {\bibfnamefont {N.}~\bibnamefont {Wiebe}}, \bibinfo
  {author} {\bibfnamefont {C.}~\bibnamefont {Gidney}}, \bibinfo {author}
  {\bibfnamefont {A.}~\bibnamefont {Aspuru-Guzik}}, \bibinfo {author}
  {\bibfnamefont {G.~K.-L.}\ \bibnamefont {Chan}},\ and\ \bibinfo {author}
  {\bibfnamefont {R.}~\bibnamefont {Babbush}},\ }\bibfield  {title} {\bibinfo
  {title} {Quantum simulation of electronic structure with linear depth and
  connectivity},\ }\href {https://doi.org/10.1103/PhysRevLett.120.110501}
  {\bibfield  {journal} {\bibinfo  {journal} {Phys. Rev. Lett.}\ }\textbf
  {\bibinfo {volume} {120}},\ \bibinfo {pages} {110501} (\bibinfo {year}
  {2018})}\BibitemShut {NoStop}%
\bibitem [{\citenamefont {Jordan}\ and\ \citenamefont
  {Wigner}(1928)}]{JordanWigner1928}%
  \BibitemOpen
  \bibfield  {author} {\bibinfo {author} {\bibfnamefont {P.}~\bibnamefont
  {Jordan}}\ and\ \bibinfo {author} {\bibfnamefont {E.}~\bibnamefont
  {Wigner}},\ }\bibfield  {title} {\bibinfo {title} {{\"U}ber das paulische
  {\"a}quivalenzverbot},\ }\href
  {https://doi.org/https://doi.org/10.1007/BF01331938} {\bibfield  {journal}
  {\bibinfo  {journal} {Zeitschrift f{\"u}r Physik}\ }\textbf {\bibinfo
  {volume} {47}},\ \bibinfo {pages} {631} (\bibinfo {year} {1928})}\BibitemShut
  {NoStop}%
\bibitem [{\citenamefont {Ortiz}\ \emph {et~al.}(2001)\citenamefont {Ortiz},
  \citenamefont {Gubernatis}, \citenamefont {Knill},\ and\ \citenamefont
  {Laflamme}}]{Ortiz2001}%
  \BibitemOpen
  \bibfield  {author} {\bibinfo {author} {\bibfnamefont {G.}~\bibnamefont
  {Ortiz}}, \bibinfo {author} {\bibfnamefont {J.~E.}\ \bibnamefont
  {Gubernatis}}, \bibinfo {author} {\bibfnamefont {E.}~\bibnamefont {Knill}},\
  and\ \bibinfo {author} {\bibfnamefont {R.}~\bibnamefont {Laflamme}},\
  }\bibfield  {title} {\bibinfo {title} {Quantum algorithms for fermionic
  simulations},\ }\href {https://doi.org/10.1103/PhysRevA.64.022319} {\bibfield
   {journal} {\bibinfo  {journal} {Phys. Rev. A}\ }\textbf {\bibinfo {volume}
  {64}},\ \bibinfo {pages} {022319} (\bibinfo {year} {2001})}\BibitemShut
  {NoStop}%
\bibitem [{\citenamefont {Dong}\ \emph {et~al.}(2021)\citenamefont {Dong},
  \citenamefont {Meng}, \citenamefont {Whaley},\ and\ \citenamefont
  {Lin}}]{Dong2021}%
  \BibitemOpen
  \bibfield  {author} {\bibinfo {author} {\bibfnamefont {Y.}~\bibnamefont
  {Dong}}, \bibinfo {author} {\bibfnamefont {X.}~\bibnamefont {Meng}}, \bibinfo
  {author} {\bibfnamefont {K.~B.}\ \bibnamefont {Whaley}},\ and\ \bibinfo
  {author} {\bibfnamefont {L.}~\bibnamefont {Lin}},\ }\bibfield  {title}
  {\bibinfo {title} {Efficient phase-factor evaluation in quantum signal
  processing},\ }\href {https://doi.org/10.1103/PhysRevA.103.042419} {\bibfield
   {journal} {\bibinfo  {journal} {Phys. Rev. A}\ }\textbf {\bibinfo {volume}
  {103}},\ \bibinfo {pages} {042419} (\bibinfo {year} {2021})}\BibitemShut
  {NoStop}%
\bibitem [{\citenamefont {Wang}\ \emph {et~al.}(2022)\citenamefont {Wang},
  \citenamefont {Dong},\ and\ \citenamefont {Lin}}]{Wang2022}%
  \BibitemOpen
  \bibfield  {author} {\bibinfo {author} {\bibfnamefont {J.}~\bibnamefont
  {Wang}}, \bibinfo {author} {\bibfnamefont {Y.}~\bibnamefont {Dong}},\ and\
  \bibinfo {author} {\bibfnamefont {L.}~\bibnamefont {Lin}},\ }\bibfield
  {title} {\bibinfo {title} {On the energy landscape of symmetric quantum
  signal processing},\ }\href {https://doi.org/10.22331/q-2022-11-03-850}
  {\bibfield  {journal} {\bibinfo  {journal} {{Quantum}}\ }\textbf {\bibinfo
  {volume} {6}},\ \bibinfo {pages} {850} (\bibinfo {year} {2022})}\BibitemShut
  {NoStop}%
\bibitem [{\citenamefont {Wille}\ \emph {et~al.}(2014)\citenamefont {Wille},
  \citenamefont {Lye},\ and\ \citenamefont {Drechsler}}]{Wille2014}%
  \BibitemOpen
  \bibfield  {author} {\bibinfo {author} {\bibfnamefont {R.}~\bibnamefont
  {Wille}}, \bibinfo {author} {\bibfnamefont {A.}~\bibnamefont {Lye}},\ and\
  \bibinfo {author} {\bibfnamefont {R.}~\bibnamefont {Drechsler}},\ }\bibfield
  {title} {\bibinfo {title} {Optimal swap gate insertion for nearest neighbor
  quantum circuits},\ }in\ \href {https://doi.org/10.1109/ASPDAC.2014.6742939}
  {\emph {\bibinfo {booktitle} {2014 19th Asia and South Pacific Design
  Automation Conference (ASP-DAC)}}}\ (\bibinfo {address} {Singapore},\
  \bibinfo {year} {2014})\ pp.\ \bibinfo {pages} {489--494}\BibitemShut
  {NoStop}%
\bibitem [{\citenamefont {Somma}\ \emph {et~al.}(2002)\citenamefont {Somma},
  \citenamefont {Ortiz}, \citenamefont {Gubernatis}, \citenamefont {Knill},\
  and\ \citenamefont {Laflamme}}]{Somma2002}%
  \BibitemOpen
  \bibfield  {author} {\bibinfo {author} {\bibfnamefont {R.}~\bibnamefont
  {Somma}}, \bibinfo {author} {\bibfnamefont {G.}~\bibnamefont {Ortiz}},
  \bibinfo {author} {\bibfnamefont {J.~E.}\ \bibnamefont {Gubernatis}},
  \bibinfo {author} {\bibfnamefont {E.}~\bibnamefont {Knill}},\ and\ \bibinfo
  {author} {\bibfnamefont {R.}~\bibnamefont {Laflamme}},\ }\bibfield  {title}
  {\bibinfo {title} {Simulating physical phenomena by quantum networks},\
  }\href {https://doi.org/10.1103/PhysRevA.65.042323} {\bibfield  {journal}
  {\bibinfo  {journal} {Phys. Rev. A}\ }\textbf {\bibinfo {volume} {65}},\
  \bibinfo {pages} {042323} (\bibinfo {year} {2002})}\BibitemShut {NoStop}%
\bibitem [{\citenamefont {McKay}\ \emph {et~al.}(2017)\citenamefont {McKay},
  \citenamefont {Wood}, \citenamefont {Sheldon}, \citenamefont {Chow},\ and\
  \citenamefont {Gambetta}}]{McKay2017}%
  \BibitemOpen
  \bibfield  {author} {\bibinfo {author} {\bibfnamefont {D.~C.}\ \bibnamefont
  {McKay}}, \bibinfo {author} {\bibfnamefont {C.~J.}\ \bibnamefont {Wood}},
  \bibinfo {author} {\bibfnamefont {S.}~\bibnamefont {Sheldon}}, \bibinfo
  {author} {\bibfnamefont {J.~M.}\ \bibnamefont {Chow}},\ and\ \bibinfo
  {author} {\bibfnamefont {J.~M.}\ \bibnamefont {Gambetta}},\ }\bibfield
  {title} {\bibinfo {title} {Efficient $z$ gates for quantum computing},\
  }\href {https://doi.org/10.1103/PhysRevA.96.022330} {\bibfield  {journal}
  {\bibinfo  {journal} {Phys. Rev. A}\ }\textbf {\bibinfo {volume} {96}},\
  \bibinfo {pages} {022330} (\bibinfo {year} {2017})}\BibitemShut {NoStop}%
\bibitem [{\citenamefont {Hatano}\ and\ \citenamefont
  {Suzuki}(2005)}]{Hatano2005}%
  \BibitemOpen
  \bibfield  {author} {\bibinfo {author} {\bibfnamefont {N.}~\bibnamefont
  {Hatano}}\ and\ \bibinfo {author} {\bibfnamefont {M.}~\bibnamefont
  {Suzuki}},\ }\bibinfo {title} {Finding exponential product formulas of higher
  orders},\ in\ \href {https://doi.org/10.1007/11526216_2} {\emph {\bibinfo
  {booktitle} {Quantum Annealing and Other Optimization Methods}}},\ \bibinfo
  {editor} {edited by\ \bibinfo {editor} {\bibfnamefont {A.}~\bibnamefont
  {Das}}\ and\ \bibinfo {editor} {\bibfnamefont {B.}~\bibnamefont
  {K.~Chakrabarti}}}\ (\bibinfo  {publisher} {Springer Berlin Heidelberg},\
  \bibinfo {address} {Berlin, Heidelberg},\ \bibinfo {year} {2005})\ pp.\
  \bibinfo {pages} {37--68}\BibitemShut {NoStop}%
\bibitem [{\citenamefont {Dong}\ and\ \citenamefont {Lin}(2022)}]{QETU2022}%
  \BibitemOpen
  \bibfield  {author} {\bibinfo {author} {\bibfnamefont {Y.}~\bibnamefont
  {Dong}}\ and\ \bibinfo {author} {\bibfnamefont {L.}~\bibnamefont {Lin}},\
  }\href {https://github.com/qsppack/QETU} {\bibinfo {title} {{QET-U}}},\
  \bibinfo {howpublished} {\url{https://github.com/qsppack/QETU}} (\bibinfo
  {year} {2022})\BibitemShut {NoStop}%
\bibitem [{\citenamefont {Cade}\ \emph {et~al.}(2020)\citenamefont {Cade},
  \citenamefont {Mineh}, \citenamefont {Montanaro},\ and\ \citenamefont
  {Stanisic}}]{Cade2020}%
  \BibitemOpen
  \bibfield  {author} {\bibinfo {author} {\bibfnamefont {C.}~\bibnamefont
  {Cade}}, \bibinfo {author} {\bibfnamefont {L.}~\bibnamefont {Mineh}},
  \bibinfo {author} {\bibfnamefont {A.}~\bibnamefont {Montanaro}},\ and\
  \bibinfo {author} {\bibfnamefont {S.}~\bibnamefont {Stanisic}},\ }\bibfield
  {title} {\bibinfo {title} {Strategies for solving the fermi-hubbard model on
  near-term quantum computers},\ }\href
  {https://doi.org/10.1103/PhysRevB.102.235122} {\bibfield  {journal} {\bibinfo
   {journal} {Phys. Rev. B}\ }\textbf {\bibinfo {volume} {102}},\ \bibinfo
  {pages} {235122} (\bibinfo {year} {2020})}\BibitemShut {NoStop}%
\bibitem [{\citenamefont {{Qiskit contributors}}(2023)}]{Qiskit}%
  \BibitemOpen
  \bibfield  {author} {\bibinfo {author} {\bibnamefont {{Qiskit
  contributors}}},\ }\href {https://doi.org/10.5281/zenodo.2573505} {\bibinfo
  {title} {Qiskit: An open-source framework for quantum computing}} (\bibinfo
  {year} {2023})\BibitemShut {NoStop}%
\bibitem [{\citenamefont {Müller}(2024)}]{GitHub}%
  \BibitemOpen
  \bibfield  {author} {\bibinfo {author} {\bibfnamefont {T.~R.}\ \bibnamefont
  {Müller}},\ }\href@noop {} {\bibinfo {title} {Ground-state preparation of
  the $2\times2$ fermi-hubbard model via {QETU}}},\ \bibinfo {howpublished}
  {\url{https://github.com/thilomueller/QETU}} (\bibinfo {year}
  {2024})\BibitemShut {NoStop}%
\bibitem [{\citenamefont {Georgopoulos}\ \emph {et~al.}(2021)\citenamefont
  {Georgopoulos}, \citenamefont {Emary},\ and\ \citenamefont
  {Zuliani}}]{Georgopoulos2021}%
  \BibitemOpen
  \bibfield  {author} {\bibinfo {author} {\bibfnamefont {K.}~\bibnamefont
  {Georgopoulos}}, \bibinfo {author} {\bibfnamefont {C.}~\bibnamefont
  {Emary}},\ and\ \bibinfo {author} {\bibfnamefont {P.}~\bibnamefont
  {Zuliani}},\ }\bibfield  {title} {\bibinfo {title} {Modeling and simulating
  the noisy behavior of near-term quantum computers},\ }\href
  {https://doi.org/10.1103/PhysRevA.104.062432} {\bibfield  {journal} {\bibinfo
   {journal} {Phys. Rev. A}\ }\textbf {\bibinfo {volume} {104}},\ \bibinfo
  {pages} {062432} (\bibinfo {year} {2021})}\BibitemShut {NoStop}%
\bibitem [{\citenamefont {Nielsen}\ and\ \citenamefont
  {Chuang}(2010)}]{Nielsen_Chuang_2010}%
  \BibitemOpen
  \bibfield  {author} {\bibinfo {author} {\bibfnamefont {M.~A.}\ \bibnamefont
  {Nielsen}}\ and\ \bibinfo {author} {\bibfnamefont {I.~L.}\ \bibnamefont
  {Chuang}},\ }\href@noop {} {\emph {\bibinfo {title} {Quantum Computation and
  Quantum Information: 10th Anniversary Edition}}}\ (\bibinfo  {publisher}
  {Cambridge University Press},\ \bibinfo {year} {2010})\ pp.\ \bibinfo {pages}
  {378--379}\BibitemShut {NoStop}%
\bibitem [{\citenamefont {Karacan}\ \emph {et~al.}(2024)\citenamefont
  {Karacan}, \citenamefont {Chen},\ and\ \citenamefont {Mendl}}]{Karacan2024}%
  \BibitemOpen
  \bibfield  {author} {\bibinfo {author} {\bibfnamefont {E.}~\bibnamefont
  {Karacan}}, \bibinfo {author} {\bibfnamefont {Y.}~\bibnamefont {Chen}},\ and\
  \bibinfo {author} {\bibfnamefont {C.~B.}\ \bibnamefont {Mendl}},\ }\href
  {https://arxiv.org/abs/2401.09091} {\bibinfo {title} {Enhancing scalability
  of quantum eigenvalue transformation of unitary matrices for ground state
  preparation through adaptive finer filtering}} (\bibinfo {year} {2024}),\
  \Eprint {https://arxiv.org/abs/2401.09091} {arXiv:2401.09091 [quant-ph]}
  \BibitemShut {NoStop}%
\bibitem [{\citenamefont {Martyn}\ \emph {et~al.}(2024)\citenamefont {Martyn},
  \citenamefont {Rossi}, \citenamefont {Cheng}, \citenamefont {Liu},\ and\
  \citenamefont {Chuang}}]{Martyn2024}%
  \BibitemOpen
  \bibfield  {author} {\bibinfo {author} {\bibfnamefont {J.~M.}\ \bibnamefont
  {Martyn}}, \bibinfo {author} {\bibfnamefont {Z.~M.}\ \bibnamefont {Rossi}},
  \bibinfo {author} {\bibfnamefont {K.~Z.}\ \bibnamefont {Cheng}}, \bibinfo
  {author} {\bibfnamefont {Y.}~\bibnamefont {Liu}},\ and\ \bibinfo {author}
  {\bibfnamefont {I.~L.}\ \bibnamefont {Chuang}},\ }\href
  {https://arxiv.org/abs/2409.19043} {\bibinfo {title} {Parallel quantum signal
  processing via polynomial factorization}} (\bibinfo {year} {2024}),\ \Eprint
  {https://arxiv.org/abs/2409.19043} {arXiv:2409.19043 [quant-ph]} \BibitemShut
  {NoStop}%
\bibitem [{\citenamefont {Kökcü}\ \emph {et~al.}(2023)\citenamefont
  {Kökcü}, \citenamefont {Camps}, \citenamefont {Oftelie}, \citenamefont
  {De~Jong}, \citenamefont {Van~Beeumen},\ and\ \citenamefont
  {Kemper}}]{Kökcü2023}%
  \BibitemOpen
  \bibfield  {author} {\bibinfo {author} {\bibfnamefont {E.}~\bibnamefont
  {Kökcü}}, \bibinfo {author} {\bibfnamefont {D.}~\bibnamefont {Camps}},
  \bibinfo {author} {\bibfnamefont {L.~B.}\ \bibnamefont {Oftelie}}, \bibinfo
  {author} {\bibfnamefont {W.~A.}\ \bibnamefont {De~Jong}}, \bibinfo {author}
  {\bibfnamefont {R.}~\bibnamefont {Van~Beeumen}},\ and\ \bibinfo {author}
  {\bibfnamefont {A.~F.}\ \bibnamefont {Kemper}},\ }\bibfield  {title}
  {\bibinfo {title} {Algebraic compression of free fermionic quantum circuits:
  Particle creation, arbitrary lattices and controlled evolution},\ }in\ \href
  {https://doi.org/10.1109/QCE57702.2023.10292} {\emph {\bibinfo {booktitle}
  {2023 IEEE International Conference on Quantum Computing and Engineering
  (QCE)}}},\ Vol.~\bibinfo {volume} {02}\ (\bibinfo {year} {2023})\ pp.\
  \bibinfo {pages} {381--382}\BibitemShut {NoStop}%
\end{thebibliography}%

\end{document}